\documentclass[journal]{IEEEtran}
\usepackage{amsmath}
\usepackage{amsfonts}
\usepackage{amssymb}
\usepackage{bm}
\usepackage{subfigure}
\usepackage{graphicx}   
\usepackage{cite}   
\usepackage{color}
\newtheorem{Lma}{Lemma}
\newtheorem{Rmk}{Remark}
\newtheorem{Def}{Definition}
\newtheorem{Tem}{Theorem}

\newtheorem{Col}{Corollary}

\newtheorem{Ass}{Assumption}

\ifCLASSINFOpdf
\else
\fi
\begin{document}
%
\title{Distributed Partial Quantum Consensus of Qubit Networks with Connected Topologies}
%
%
%

\author{
	Xin~Jin,~\IEEEmembership{Member,~IEEE}, 	
	Zhu~Cao,
	 Yang~Tang,~\IEEEmembership{Fellow,~IEEE}, 
	J\"{u}rgen Kurths
\thanks{}
\thanks{
	This work was supported in part by the National Natural Science Foundation of China under
	Grant 62233005 and Grant 12105105, 
	the Young Elite Scientist Sponsorship Program by Cast under Grant
	YESS20220198, the Natural Science Foundation of Shanghai
	under Grant 21ZR1415800, the Shanghai Sailing Program under Grant 23YF1409600 and Grant 21YF1409800, the Fundamental Research Funds for the Central Universities under Grant 2024SMECP06, and the startup fund from East
	China University of Science and Technology under Grant
	YH0142214.
	\textit{(Corresponding authors: Zhu Cao, Yang Tang.)}}
	\thanks{
		Xin Jin is with the Key Laboratory of
		Smart Manufacturing in Energy Chemical Process, Ministry of Education,
		East China University of Science and Technology, Shanghai, 200237, 
		China, and with the Research Institute of Intelligent Complex Systems, Fudan University, Shanghai, 200433, China.}
\thanks{
	Zhu Cao and Yang Tang are with the Key Laboratory of Smart Manufacturing in Energy Chemical Process, Ministry of Education, East China University of Science and Technology, Shanghai 200237, China (e-mail: caozhu@ecust.edu.cn; tangtany@gmail.com).}
\thanks{J\"{u}rgen Kurths is with the Potsdam Institute for Climate Impact Research,
	14473 Potsdam, Germany, and Institute of Physics, Humboldt University of Berlin, 12489
	Berlin, Germany.
}
}
%
%

\markboth{IEEE Transactions on Cybernetics, ~Vol.~**, No.~*, ***~2024}%
{Shell \MakeLowercase{\textit{et al.}}: Bare Demo of IEEEtran.cls for IEEE Journals}
%



\maketitle

\begin{abstract}
 In this paper, {\color{black}we consider the partial quantum consensus problem of a qubit network in a distributed view. 
 	The local quantum operation is designed based on the Hamiltonian by using the local information of each quantum system in a network of qubits.} 
We construct the unitary transformation for each quantum system to achieve the partial quantum consensus, i.e., the directions of the quantum states in the Bloch ball will reach an agreement. 
{\color{black}A simple case of two-qubit quantum systems is considered first, and a minimum completing time of reaching partial consensus is obtained based on the geometric configuration of each qubit. }
Furthermore, we extend the approaches to deal with the more general $N$-qubit networks.
{\color{black}Two partial quantum consensus protocols, based on the Lyapunov method for chain graphs and the geometry method for connected graphs, are proposed.}
{\color{black}The geometry method can be utilized to deal with more general connected graphs, while for the Lyapunov method, the global consensus can be obtained. }
{\color{black}The numerical simulation over a qubit network is demonstrated to verify the validity and the effectiveness of the theoretical results.}
\end{abstract}

\begin{IEEEkeywords}
quantum system, consensus, qubit network, connected graphs
\end{IEEEkeywords}

%
\IEEEpeerreviewmaketitle

\section{Introduction}
%
%
%
%
{\color{black}Consensus
problems have been studied in the field of distributed computing in the early stage, and have wide applications in distributed control and optimization of networked systems \cite{2023_Chaos_X.Jin,2020_Auto_X.Jin,2022_Auto_X.Jin,2021_X.Jin_TAC}.
{\color{black}In the past decade, 
 the quantum network has been an emerging concept both in  theoretical studies and physical implementations of distributed quantum computing\cite{2023_Quantum_TCYB}, secure communication \cite{2017_Quantum_Secure}, and quantum machine learning \cite{2019_TCYB_donglearning,comments1,comments2,comments3,comments4,comments5}.}}
{\color{black}In a quantum network, quantum information is generated, processed,  and stored locally in quantum nodes, which are linked through communication channels \cite{2022_ARC_quantum}.}

Seeking the quantum consensus protocol, which ensures the agreement in a quantum network, has been a fundamental step in the quantum applications \cite{2023_TCNS_quantum}.
{\color{black}The pioneering work in quantum consensus has been made in \cite{Mazzarella2015}, which defines four types of quantum consensus states}. 
Based on the classical gossip algorithm, a quantum gossip consensus protocol is proposed, which can achieve the symmetric state consensus. 
Two improved gossip-like consensus algorithms are proposed in \cite{Ticozzi2016} to reach a purer output state and guarantee the symmetric state consensus simultaneously.
 {\color{black}The connection between quantum consensus and classical consensus is initially established by introducing an induced graph in\cite{Shi2016,2015_ACC}. 
 	Here, quantum dynamics are modeled using master equations in Lindblad form, and interactions among quantum subsystems are described using the swapping operator.}
This research gives necessary and sufficient conditions for exponential and asymptotic (symmetric-state) consensus convergence for undirected connected graphs.
{\color{black} Furthermore, in \cite{Shi2017}, it considers the average consensus of quantum networks under the balanced directed graph. The result is extended to switching interactions based on the properties of cut-balanced graphs.} 
{\color{black}However, the induced graph in \cite{Shi2016,Shi2017} is modeled by a  $2^{N}\times 2^{N}$ dimensional Laplacian matrix, of which dimension exponentially increases with the number of nodes in quantum networks. }
{\color{black}The exponential increase in the dimension of the Laplacian matrix of the induced graph will induce a heavy computation burden as the number of qubits is large.}
The converge rate of the consensus algorithm based on quantum consensus master equations (QCMEs) is investigated in \cite{2016_AuCC,Jafarizadeh2016,Jafarizadeh2017}, and the result shows the optimal convergence rate of quantum consensus (the second smallest eigenvalue of the Laplacian matrix) can be obtained from the Laplacian matrix of the partitions, which is not directly related to the Laplacian matrix of the whole induced graph.
{\color{black} In \cite{2023_TCNS_quantum}, an open quantum network of qubits is considered with sequential measurements described by a probabilistic Boolean network. Some structure characteristics for induced Boolean networks are established from the relationship between the measurement bias and  the master equation. 
}

It should be noted that the above works\cite{Ticozzi2016,Shi2016,Shi2017,Jafarizadeh2016,Jafarizadeh2017,2023_TCNS_quantum} consider the qubit network as a whole instead of from the perspective of a multi-agent system.
More specifically, the consensus protocol was proposed based on a global operator performed on a tensor product state of the quantum network, whose dimension is exponentially increasing with the network scale \cite{Ticozzi2016,Shi2016,Shi2017,Jafarizadeh2016,Jafarizadeh2017}.
This is different from the distributed consensus protocol in multi-agent systems, which only governs the local behavior of each agent. 
In other words, the quantum consensus protocol in \cite{Ticozzi2016,Shi2016,Shi2017,Jafarizadeh2016,Jafarizadeh2017} cannot be used in a fully distributed way.  
{\color{black}This makes the results less feasible in a large-scale qubit network such as the quantum internet \cite{2008_nature}. }

{\color{black}On the other hand, the dynamics of the quantum system in \cite{Ticozzi2016,Shi2016,Shi2017,Jafarizadeh2016,Jafarizadeh2017} assume that the Hamiltonian is zero, 
which means that the local operation is not considered for each quantum system. }
{\color{black} Designing a Hamiltonian in the quantum feedback control is an effective approach in quantum control systems \cite{2023_TAC_quantum_control,2023_TAC_quantum_control_Y.Wang,2022_TQE_quantum,2023_PRL_quantum_control}.}
The controllability of quantum systems is defined in \cite{2022_ARC_quantum}. 
The main idea is to design the Hamiltonian driving the quantum states to the desired ones.
Lyapunov functions are powerful tools for feedback control design and asymptotic convergence analysis in classical control problems.
{\color{black} However, in quantum control systems, the measurement action usually destroys the state being measured, which brings difficulties in designing the feedback control based on Lyapunov functions directly \cite{Kuang2008,2021_TCYB_K.Sen,2017_TCYB_K.Sen}. }
{\color{black} A feedback design and open-loop control strategy has been proposed in the existing research \cite{Kuang2019}.
{\color{black} Nevertheless, implementing this strategy in complex network systems, such as multi-agent systems, poses significant challenges.}
{\color{black}This is due to the fact that the control action of each agent is correlated with the interaction of neighbors, which cannot be determined on the computer previously. }

{\color{black}
Motivated by the above discussions, we exploit the quantum consensus problem from a multi-agent system perspective.
{\color{black} The main difficulty of this approach is to construct the local operation which is only based on the local information of each quantum system. 
	In this work, we assume that
each qubit will interact with its neighboring qubits through weak measurements.
Such a measurement can be made to determine the density matrix of quantum systems with high accuracy, while slightly disturbing the quantum state itself \cite{2011_Nature_Lundeen}.
The transition between the pure quantum states can be completed by a unitary transformation based on the local operator. }
Due to the entanglement of quantum states, it is difficult to define the quantum consensus based on the classical counterpart.  
Hence, by decoupling quantum states into the pure states and the maximum mixed states based on the convexity property of the density matrix, we 
consider reaching partial quantum consensus based on the local operation.
 The main contributions of this work are summarized as follows:
 

{1) A minimum time of completing the unitary transformation for a two-qubit system is obtained based on {the geometric configuration} of the quantum state, and the corresponding Hamiltonian is constructed for each qubit. 
Compared with the minimum time of the unitary transformation for two-qubit systems in the recent work \cite{Dong2010}, we only need to perform a single rotation which is more efficient.}

{\color{black}
2) 
A partial quantum consensus protocol for $N$-qubit networks with chain topologies is proposed by means of the Lyapunov analysis. The Lyapunov function is designed based on the Euclidean distance between two qubits.
Compared with the quantum steering control based on Lyapunov approaches in \cite{Kuang2008}, we propose a quantum consensus protocol of {a qubit network}. The main difficulty lies in deriving the Hamiltonian
operator of each qubit since it is both affected
by quantum mechanics and network topologies.}

{3) 
A geometric approach is proposed which
 builds a connection between the partial quantum consensus
 and consensus on the 2-sphere. }Then, 
the partial quantum consensus protocol with general connected graphs is constructed based on the perpendicular relationship between the tangent direction and the rotation axis
of a point on the Bloch ball. 
{Compared with the quantum consensus protocols in \cite{Ticozzi2016,Shi2016,Shi2017,Jafarizadeh2016,Jafarizadeh2017}, we consider the quantum consensus problem from a multi-agent perspective and design a distributed consensus protocol by only using the neighbors' quantum state.}}

 The remaining part of this paper is organized as follows. 
 In Section II, the preliminary knowledge and problem formulation are presented. 
 The partial quantum consensus of two-qubit systems is investigated in Section III.
 The main results of partial quantum consensus of $N$-qubit network are shown
 in Section IV. Section V gives the numerical simulations 
 and the conclusion is drawn in Section \uppercase\expandafter{\romannumeral6}
  finally.

 \text{Notations}: 
 $\mathbb{R}^{N}$ and $\mathbb{R}^{{N}\times {N}}$ represent the Euclidean vector space and real matrix space, where $N$ is a positive integer number. $\mathbb{N}^{+}$ denotes the positive integer number.  $\mathbb{R}^{+}$ represents the positive real number.
 $\mathbb{C}^{N}$ represents the $N$ dimensional complex vector space.
 $\mathbf{1}_{N}=[1,...,1]^{\top}\in \mathbb{R}^{N}$.
 For a vector $\mathbf{x}=[x_1,...,x_N]^{\top}\in \mathbb{R}^{N}$, $\|\mathbf{{x}}\|$ denotes the Euclidean norm, which is defined as $\|\mathbf{x}\|=\sqrt{x_{1}^{2}+x_{2}^{2}+...+x_{N}^{2}}$.
 $\operatorname{tr}(\mathbf{A})$ represents the trace of matrix $\mathbf{A}$.
 $\times$ represents the cross product of two vectors in Euclidean space.
  $\mathbb{S}^{2}$ represents the 2-sphere defined as $\mathbb{S}^{2}=\{\mathbf{x}\in \mathbb{R}^{3}: \mathbf{x}^{{\top}}\mathbf{x}=1 \}$.
 A three-dimensional special orthogonal group is denoted as $\mathbb{SO}(3)=\{\mathbf{R}\in \mathbb{R}^{3 \times 3}: \mathbf{R}^{{\top}}\mathbf{R}=\mathbf{I}_{3}, \text{det}(\mathbf{R})=1\}$.
 The set of three-dimensional skew symmetric matrices is denoted as $\mathfrak{so}(3)=\{\mathbf{S}\in \mathbb{R}^{3 \times 3}: \mathbf{S}^{\top}=-\mathbf{S}\}$.
The two dimensional special unitary group is denoted as $\mathbb{SU}(2)=\{\mathbf{U}\in \mathbb{C}^{2 \times 2}: \mathbf{U}^{{\dagger}}\mathbf{U}=\mathbf{I}_{2}, \text{det}(\mathbf{U})=1\}$.
$\text{Skew}(\cdot)$ represents a mapping from a vector $\mathbf{x}=[x_{1},x_{2},x_{3}]^{\top}\in\mathbb{R}^{3}$ to a skew symmetric matrix $\mathbf{X}=
\left[\begin{matrix}
0&x_{3}& -x_{2}\\
-x_{3}&0& x_{1}\\
x_{2}&-x_{1}&0\\
\end{matrix}\right].
$

\section{Preliminaries and Problem Formulation}
\subsection{Quantum systems}
1) \textit{Pure states and mixed states:}
In quantum mechanics, a quantum state is represented as a unit vector $|\psi\rangle $ in a Hilbert space $\mathcal{H}$, where $|\cdot \rangle$ is Dirac's notation (called a \textit{ket}). 
Its dual vector is denoted as $|\psi \rangle^{\dagger}=\langle \psi| $ (called a \textit{bra}). 
The inner product $\langle \psi | \phi \rangle$ which maps a pair of vectors to a complex number $C$. Particularly, we have 
 $\langle \psi | \psi \rangle=1$. 
A quantum state that can be represented by a unit vector $|\psi \rangle$ is called a pure state. 
In an open quantum system, a state could be a mixture of a number of pure states which is called the mixed state.
Supposed that a quantum system is in a state $| \psi_{i} \rangle$ with probability $p_{i}$ such that $\sum_{i} p_{i}=1$ where $0<p_{i}\leq 1$, we 
can represent it by using the \textit{density matrix} 
\begin{align}
\bm{\rho}=\sum_{i}p_{i}| \psi_{i}\rangle \langle \psi_{i}|.
\end{align}
Mathematically, a density matrix is a self-adjoint operator $\bm{\rho}=\bm{\rho}^{\dagger}$ and satisfies $\text{tr}(\bm{\rho})=1$.
Generally, a state in quantum systems can be represented by a density operator either it is a pure state or a mixed state. 
For a pure state,  the density matrix can be expressed as  $\bm{\rho}=| \psi \rangle \langle \psi |$, where $\text{tr}(\bm{\rho}^{2})=1$.
Note that if the state is not pure, $\text{tr}(\bm{\rho}^{2})<1$.

Due to the convex property of the density matrix \cite{Quantumlecture2015}, we can express $\bm{\rho}$ as a convex combination of other two states
\begin{align}\label{convex}
  \bm{\rho}=P\frac{1}{2}\mathbf{I}_{2}+(1-P)\bm{\rho}',
\end{align}
where $P$ is a positive real number, $\mathbf{I}_{2}$ is the maximally mixed state and $\bm{\rho}'$ is the pure state.

2) \textit{Qubits and Quantum networks:} 
 The smallest unit of quantum information is a qubit in a two-dimensional Hilbert space $\mathcal{H}_{2}$. Suppose that the $|0\rangle$ and $|1\rangle$ are two-dimensional orthonormal basis. A qubit is expressed as 
\begin{align}\label{qubit}
\mathbf{q}=a|0\rangle +b|1\rangle,
\end{align} 
where $a,b$ are two complex numbers such that $|a|^{2}+|b|^{2}=1$.

In this work, we consider a quantum network with $N$ qubits. 
It is worth mentioning that the qubit in (\ref{qubit}) can be interpreted as the spin-$\frac{1}{2}$ object. 
A spin-$\frac{1}{2}$ representation of a qubit expressed in the world frame $\mathcal{W}$ is 
\begin{align}\label{spin}
|\psi_{i}(\theta_{i}, \phi_{i})\rangle^{\mathcal{W}} = 
\left(\begin{matrix}
e^{-i\frac{\phi_{i}}{2}}\cos\frac{\theta_{i}}{2}\\
e^{i\frac{\phi_{i}}{2}}\sin\frac{\theta_{i}}{2}
\end{matrix}\right),
\end{align}
where the polar angle $\theta_{i}$ and the azimuthal angle $\phi_{i}$ describe the orientation of the spin in a three-dimensional space in Fig. \ref{2qubits}A.

Each qubit related to a local body frame $\mathcal{B}_{i}$ is denoted as $|\psi_{i}(\theta_{i}, \phi_{i})\rangle^{\mathcal{B}_{i}}$. The relationship between qubits in different frames can be expressed as 
$
  |\psi_{i}(\theta_{i}, \phi_{i})\rangle^{\mathcal{W}} = \mathbf{U}_{i}|\psi_{i}(\theta_{i}, \phi_{i})\rangle^{\mathcal{B}_i},
$
where $\mathbf{U}_{i}\in \mathbb{SU}(2)$ is a unitary matrix describing the rotation from the body frame $\mathcal{B}_{i}$ to the world frame $\mathcal{W}$.
More generally, 
$
\bm{\rho}_{i}^{\mathcal{W}} = \mathbf{U}_{i}\bm{\rho}_{i}^{\mathcal{B}_i}\mathbf{U}_{i}^{\dagger}.
$
For the presentation clarity, we omit the $\mathcal{W}$ if the state is expressed in the world frame in the following.

{\color{black}
The evolution of the spin-$\frac{1}{2}$ object in time is governed by the $Schr\ddot{o}dinger\; equation$
\begin{align}\label{schr}
	\frac{\text{d}}{\text{dt}}|\psi_{i}(t) \rangle=-i\mathbf{H}_{i}|\psi_{i}(t) \rangle,
\end{align}
where $\mathbf{H}_{i}$ is a constant self-adjoint operator called the Hamiltonian. Hence, we have 
\begin{align}\label{spin_2}
	|\psi_{i}(t)\rangle = \mathbf{U}_{i}(t)|\psi_{i}(0)\rangle,
\end{align}
where $\mathbf{U}_{i}(t)=e^{-it\mathbf{H}_{i}}$ is a unitary transform. }
  \begin{figure}[htb]	
	\centering	
	{
		\includegraphics[scale=0.2]{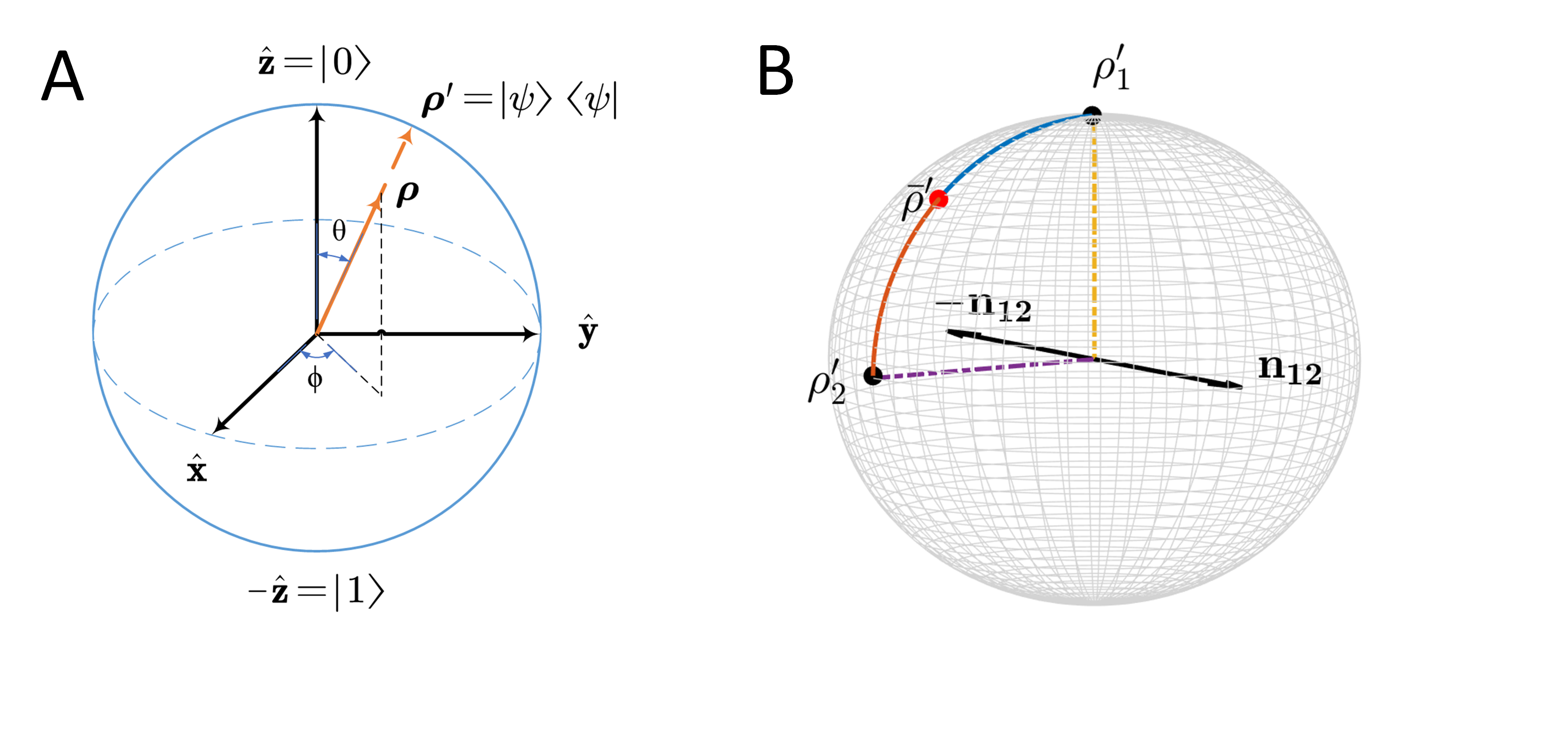}
	
	}
	\caption{A) A density matrix can be seen as a point on the Bloch ball and a pure state lies on the sphere of the Bloch ball. B) The trajectories of two qubits evolving on the Bloch ball.}
		\label{2qubits}
\end{figure}
\subsection{Graph Theory}
The interaction among $N$ qubits in a network can be considered as a graph $\mathcal{G}=\{\mathcal{V},\mathcal{E}\}$, where $\mathcal{V}=\{1,...,N\}$ denotes the 
node set and $\mathcal{E} \subseteq\mathcal{V}\times \mathcal{V}$ denotes the edge set.
A directed edge $(i,j)\in \mathcal{E}$ means that there exists a communication link from node $i$ to node $j$. 
A graph is called undirected if the edge $(i,j)\in \mathcal{E}, i,j \in \mathcal{V}$ then $(j,i)\in \mathcal{E}$, and it is connected if there exists at least one path for any two nodes in the graph. 
An adjacency matrix associated with the graph $\mathcal{G}$ is defined as $\mathbf{A}=\{a_{ij}\}\in\mathbb{R}^{N \times N}, i,j=1,...,N$, where $a_{ij}>0$ if  $\{j,i\}\in \mathcal{E}$, and $a_{ij}=0$ otherwise.
The node $j$ is node $i$'s neighbor if there is an edge connected from node $j$ to node $i$. A neighboring set  $\mathcal{N}_i=\{j\in\mathcal{V}, (j,i)\in\mathcal{E}\}$ consists of all the neighbors of node $i$.

\subsection{Problem Formulation}
In an $N$-qubit network, we express the state of each qubit as a density matrix $\bm{\rho}_{i}, i\in \mathcal{V}$.
Based on the property of the density matrix, we can present each qubits' state in the general form 
\begin{align}{\label{density}}
\bm{\rho}_{i}=\frac{1}{2}\Big(\mathbf{I}_{2}+\mathbf{p}_{i}\cdot\vec{\bm{\sigma}}\Big),
\end{align}
where $\mathbf{p}_{i}=[p_{1},p_{2},p_{3}]^{\top}\in \mathbb{R}^{3}$ is a vector which satisfies $0 \leq\|\mathbf{p}_{i}\| \leq1$ and 
$\mathbf{p}_{i}\cdot\vec{\bm{\sigma}}=p_{1}\bm{\sigma}_{1}+p_{2}\bm{\sigma}_{2}+p_{3}\bm{\sigma}_{3}$,
$\bm{\sigma}_{i},i=1,2,3$ are Pauli matrices defined as 
\begin{align}
	\bm{\sigma}_{1}=\left[
\begin{matrix}
	0&1\\
	1&0\\
\end{matrix}\right],
	\bm{\sigma}_{2}=\left[
\begin{matrix}
	0&-i\\
	i&0\\
\end{matrix}\right],
	\bm{\sigma}_{3}=\left[
\begin{matrix}
	1&0\\
	0&-1\\
\end{matrix}\right]. \nonumber
\end{align}
Therefore, there is a one-to-one mapping  between a density matrix and a point in a three-dimensional unit ball (called Bloch ball) presented in Fig. \ref{2qubits}A.

{\color{black}Note that the quantum state for each system $\bm{\rho}_{i}$ can be a mixed state. 
According to (\ref{convex}), we can separate the state into a maximally mixed state and a pure state.}
Based on (\ref{spin}), we can rewrite the pure state in the following form,
\begin{align}{\label{puredensity}}
  \bm{\rho}'_{i}(\mathbf{u}_{i})=\frac{1}{2}\Big(\mathbf{I}_{2}+\mathbf{u}_{i}\cdot\vec{\bm{\sigma}}\Big),
\end{align}
where $\mathbf{u}_{i}=\left( \sin\theta_{i} \cos\phi_{i}, \sin\theta_{i} \sin\phi_{i}, \cos \theta_{i} \right)$ is a unit vector. 
From (\ref{puredensity}), we know that each pure state $ \bm{\rho}'_{i}(\mathbf{u}_{i})$ lies on the sphere of the Bloch ball shown in Fig. \ref{2qubits}A.
Thus, a pure state $\bm{\rho}'_{i}=|\psi_{i} \rangle  \langle \psi_{i}|$ can be transferred to another pure state $\bm{\rho}'_{j}=|\psi_{j} \rangle  \langle \psi_{j}|$ through a unitary transformation $\mathbf{U}_{ij}$, i.e.,
$
  \bm{\rho}'_{j}=\mathbf{U}_{ij}\bm{\rho}'_{i}\mathbf{U}^{\dagger}_{ij}
$,
and it is also possible to construct a unitary transformation for each quantum system $\bm{\rho}_{i}$ to drive the part of the pure state of each quantum system to partial quantum consensus. 
{\color{black}More specifically, assuming that the unitary transformation for the pure state $\rho_{i}^{\prime}$ in the part of the mixed state $\rho_{i}$ is defined as $\mathbf{U}_{i}$, i.e., $\mathbf{U}_{i}\bm{\rho}'_{i}\mathbf{U}_{i}^{\dagger}=\bar{\bm{\rho}}_{i}^{\prime}$, where $\bar{\bm{\rho}}_{i}^{\prime}$ denotes the consensus pure state. Then, by utilizing the unitary transformation to the mixed state $\rho_{i}$, we have 
	\begin{align}{\label{partial}}
		\mathbf{U}_{i}\bm{\rho}_{i}\mathbf{U}_{i}^{\dagger}=&\mathbf{U}_{i}\left(P\frac{1}{2}\mathbf{I}_{2}+(1-P)\bm{\rho}'_{i}\right)\mathbf{U}_{i}^{\dagger}\nonumber\\
		=&P\frac{1}{2}\mathbf{I}_{2}+(1-P)\bar{\bm{\rho}}'_{i}.
	\end{align}
	From (\ref{partial}), we know that based on the unitary transformation $\mathbf{U}_{i}$, the pure state $\bm{\rho}_{i}^{\prime}$ achieves the consensus. This allows us to only design the unitary transformation for the pure state $\bm{\rho}_{i}^{\prime}$ in the part of the mixed state $\bm{\rho}_{i}$ to achieve the partial quantum consensus as defined in Definition 1.}
The pure state $\bm{\rho}'_{i}$ can be calculated as 
$
	\bm{\rho}'_{i}=\frac{1}{2}\Big(\mathbf{I}_{2}+\frac{\mathbf{p}_{i}}{\|\mathbf{p}_{i}\|}\cdot\vec{\bm{\sigma}}\Big),\nonumber
$
where $\mathbf{p}_{i}$ is defined in (\ref{density}).
The relationship between the quantum state $\bm{\rho}_{i}$ and its related pure state $\bm{\rho}'_{i}$ is shown in Fig. \ref{2qubits}A.
\begin{Def}\cite{Shi2016,Jafarizadeh2017}
Supposed that a quantum network of $N$-qubit which is governed by (\ref{schr}), the partial quantum consensus is achieved if all the quantum states point in the same direction on the Bloch ball, i.e.,
\begin{align}
  \bm{\rho}'_{1}=  \bm{\rho}'_{2}=...=  \bm{\rho}'_{N}=\bar{\bm{\rho}}^{\prime}.
\end{align} 
\end{Def}
{\color{black}where $\bar{\bm{\rho}}^{\prime}$ is the consensus state of all qubits.}

{\color{black}
In this problem, we consider a multi-agent system consisting of $N$ qubits, where the state of each qubit is represented by a density matrix $\bm{\rho}_{i}\in \mathbb{C}^{2 \times 2}$, and the quantum state of the whole qubit network is a product state. We assume that the local interaction between qubits is made through the weak measurement \cite{1999_PRA_Quantum,2011_Nature_Lundeen}.}
For the pure states, the dynamics is governed by the $Schr\ddot{o}dinger\; equation$ (\ref{schr}).
The main contribution of this work is to design the interactive Hamiltonian $ \mathbf{H}_{i}, \;i=1,...,N$, which is also constructing the unitary transformation of each qubit to complete the partial quantum consensus.

In the interactive Hamiltonian of each qubit, the local information, which is its own quantum state  $\bm{\rho}^{\prime}_{i}$ and its neighbors' quantum states $\bm{\rho}^{\prime}_{j},j\in \mathcal{N}_{i}$, will be used. 
To obtain the neighbors' quantum states, we assume that each qubit can make a weak measurement to its neighbors, while not destroying its neighbors' state.
It is worth mentioning that this Hamiltonian $\mathbf{H}_{i}$ is similar to the protocol in multi-agent systems, which specifies the information that each qubit interacts with its neighbors. Hereafter, we will call the interactive Hamiltonian $\mathbf{H}_{i}, \;i=1,...,N$, in {\color{black}a qubit network} as the \emph{quantum protocol}.  
\begin{Rmk}
         {\color{black}The quantum consensus algorithm is proposed based on QCMEs in \cite{2015_ACC,2016_AuCC,Shi2016,Shi2017}. Based on a swapping operator, the reduced state synchronization is achieved.
         However, this swapping operator is a global operation for the tensor product state of qubit networks. }
         Note that this operation is not suitable in the scenarios when two nodes are located in different places which are far away from each other such as quantum internet. 
         To seek a fully distributed quantum consensus algorithm, we aim to propose a local operator based on the local information for each qubit to reach consensus.
         From this perspective, we design the local interactive Hamiltonian of each qubit. 
         In order to reach quantum consensus, the state information of neighboring qubits should be included in the Hamiltonian design of each subsystem.  
         To facilitate the theoretical analysis, we use
         the weak measurement assumption in this work, which aims to achieve the local interaction in qubit networks while preserving quantum coherence.
         Specifically, qubit $i$ can measure the density matrix $\bm{\bm{\rho}}_{j}$ of qubit $j$ while not destroying the state $\bm{\bm{\rho}}_{j}$ through the weak measurement. The details on weak measurements are referred to \cite{1999_PRA_Quantum}.   
\end{Rmk}
\begin{figure}[htb]	
	
	\centering
	{
		\includegraphics[scale=0.20]{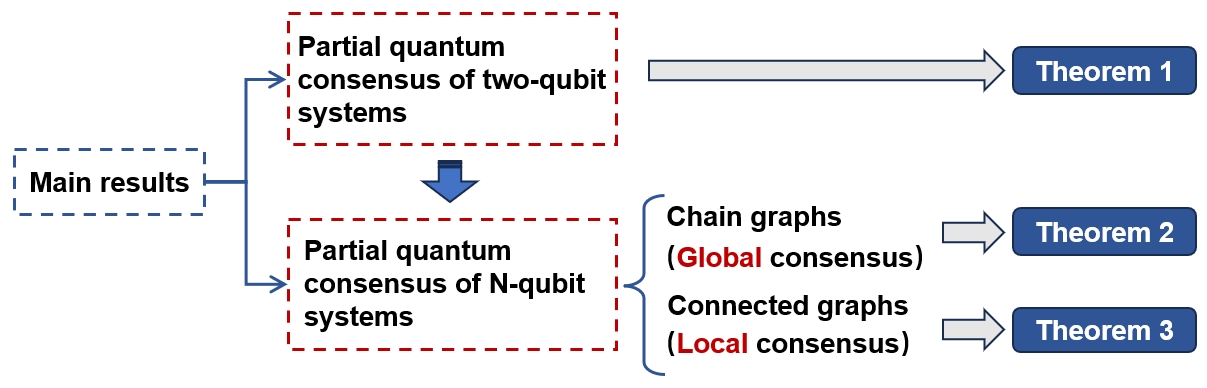}
	}
	\caption{The relationships among main results.}
	\label{result}
\end{figure} 
\section{Partial quantum consensus for two-qubit systems}
In this section, we firstly consider the simple case where quantum networks only have two qubits. 
The settling time is one of the important criteria in the control task. 
In this part, we firstly give an explicit expression for the minimum time that two qubits needed to complete the unitary transformation.  
This problem can be seen as a stepping stone for the partial quantum consensus protocol design for the quantum networks with more qubits.  A diagram of the main theoretical results is shown in Fig. \ref{result}.

For two-qubit systems, the quantum consensus in Definition 1 reduces to 
$|\psi_{i}(\theta_{i}, \phi_{i})\rangle =|\psi_{j}(\theta_{j}, \phi_{j})\rangle, i=1, j=2$.
Let $\mathbf{s}_{i}(\theta_{i}, \phi_{i})$ and $\mathbf{s}_{j}(\theta_{j}, \phi_{j})$ be the corresponding two points of the qubits in the spherical coordinates.
A direct idea is to construct a single unitary transformation  $\mathbf{U}_{ij}$ from  $\mathbf{s}_{i}(\theta_{i}, \phi_{i})$ to  $\mathbf{s}_{j}(\theta_{j}, \phi_{j})$.  
A unitary transformation $\mathbf{U}_{ij}(t)\in \mathbb{SU}(2)$ can represent a rotation $\theta_{ij}$ of the spin around the axis $\mathbf{n}_{ij}$ in the following form,
\begin{align}\label{unitary}
\mathbf{U}_{ij}(\mathbf{n}_{ij},\theta_{ij})=e^{-i\frac{\theta_{ij}}{2}\mathbf{n}_{ij}\cdot \vec{\bm{\sigma}}},
\end{align}
where $\vec{\bm{\sigma}}$ are the Pauli matrices. 
We can compare (\ref{unitary}) with (\ref{spin_2}). The Hamiltonian can be designed as $\mathbf{H}_{i}=\mathbf{n}_{ij}\cdot \vec{\bm{\sigma}}$, then the completing time of the unitary transformation will be ${\theta_{ij}}$. 
The trajectories of two qubits under the unitary transformation are shown in Fig. \ref{2qubits}B.
The rotation axis $\mathbf{n}_{12}$ is designed based on the cross product of both direction vectors of $\bm{\rho}^{\prime}_{1}$ and  $\bm{\rho}^{\prime}_{2}$. 
It shows that the two pure states can rotate to the same point (red marker) on the Bloch ball.
\begin{Tem}
	Suppose that a two-qubit system governed by (\ref{schr}). There exists a single unitary transformation $\mathbf{U}_{ij}(\mathbf{n}_{ij},\theta_{ij})$ such that $\mathbf{U}_{ij}(\mathbf{n}_{ij},\theta_{ij})|\psi_{i}(\theta_{i}, \phi_{i})\rangle =\mathbf{U}_{ij}^{\dagger}(\mathbf{n}_{ij},\theta_{ij})|\psi_{j}(\theta_{j}, \phi_{j})\rangle$. Furthermore, the minimum time for completing the rotation is 
	\begin{align}{\label{Tmin}}
	   T=\frac{1}{2}\arccos (\mathbf{s}_{i}\cdot \mathbf{s}_{j}),
	\end{align}
	where $\mathbf{s}_{i}(\theta_{i}, \phi_{i})=\left(  \sin\theta_{i} \cos\phi_{i}, \sin\theta_{i} \sin\phi_{i}, \cos \theta_{i}   \right)$ and $\mathbf{s}_{j}(\theta_{j}, \phi_{j})=\left(  \sin\theta_{j} \cos\phi_{j}, \sin\theta_{j} \sin\phi_{j}, \cos \theta_{j}   \right).$ 
\end{Tem}
\begin{IEEEproof}
	Since we know that there exists a one-to-one mapping from the qubit $|\psi_{i} \rangle$ to the vector $\mathbf{s}_{i}(\theta_{i}, \phi_{i})$ in the spherical coordinate, 
	the two vectors $\mathbf{s}_{i}(\theta_{i}, \phi_{i})$ and $\mathbf{s}_{j}(\theta_{j}, \phi_{j})$ can form a plane $\mathbb{P}$. 
	Hence, we can construct a rotation along the direction of the normal vector on the plane $\mathbb{P}$ and the rotation angle is $\theta_{ij}$.
	The unitary transformation of the rotation can be expressed as 	$\mathbf{U}_{ij}(\mathbf{n}_{ij},\theta_{ij})=e^{-i\frac{\theta_{ij}}{2}\mathbf{n}_{ij}\cdot \vec{\bm{\sigma}}}$, where
	\begin{align}
\mathbf{n}_{ij}=&\;\mathbf{s}_{i}(\theta_{i}, \phi_{i})\times \mathbf{s}_{j}(\theta_{j}, \phi_{j})\nonumber \\ 
=&\;\left[\begin{matrix}
\sin \theta_{i}\cos \theta_{j} \sin \phi_{i}  -  \cos \theta_{i}\sin \theta_{j} \sin \phi_{i} \\
\cos \theta_{i} \sin \theta_{j} \cos \phi_{j} -  \sin \theta_{i} \cos \theta_{j} \cos \phi_{i}\\
\sin \theta_{i} \sin \theta_{j} \cos \phi_{i} \sin \phi_{j} - \sin \theta_{i} \sin \theta_{j} \sin \phi_{i} \cos \phi_{j}
\end{matrix}\right].\nonumber
	\end{align}
	The Hamiltonian for qubit $i$ and qubit $j$ can be determined as $\mathbf{n}_{ij}$ and $-\mathbf{n}_{ij}$, respectively. 
	Thus, the minimum time for completing the unitary transformation is $T=\frac{1}{2}\arccos (\mathbf{s}_{i}(\theta_{i}, \phi_{i})\cdot \mathbf{s}_{j}(\theta_{j}, \phi_{j}))$, which completes the proof.  
\end{IEEEproof}
{\color{black}
\begin{Rmk}
	The minimum time is similar to the concept of the finite-time, which both achieve the control task in a settling time \cite{2021_TCYB_Q.L.Han}. The difference is that the finite-time controller is designed based on an analytic solution of differential equations while the minimum completing time in this work is achieved by constructing a unitary transformation based on the geometric configuration of qubits on the Bloch ball.
	A similar problem is considered in \cite{Dong2010}, which derives an analytical expression of the minimum time for completing the unitary 
	transformation from the initial pure states to the target pure states. However, the designing unitary transformation is completed by the $x-z-x$ rotations, and the minimum time is based on the assumption with the unbound control. In this work, we only design the Hamiltonian with a one-step unitary transformation, which is more straightforward and efficient. 
\end{Rmk}}

\section{Partial quantum consensus for $N$-qubit networks}
It is obvious that a two-qubit system is quite particular.
In this section, we consider the more general case where the number of qubits is $N$ by extending the approaches in the two-qubit case.
Two different topology cases are investigated, respectively. 
{For the chain graph, the Hamiltonian for each quantum system is designed by the Lyapunov approach and the global convergence result is obtained. However, it is difficult to generalize the result to a connected graph based on the Lyapunov approach.
{\color{black}Thus, for the general connected graph, we build the relationship of this problem with the consensus on 2-sphere and design the Hamiltonian based on the geometric configuration of the quantum state on the Bloch ball while the local partial quantum consensus is achieved. } }
{\color{black}\subsection{Lyapunov-based quantum protocol design}}
{\color{black}The chain graph} is firstly considered in the following. 
This configuration of quantum networks can be seen in the physical applications such as the linear ion trap, which is used as a component of a trapped ion quantum computer.  
\begin{Ass}
	The topology of an $N$-qubit network is a chain graph. 
\end{Ass}

For each qubit, the quantum protocol is given as
\begin{align}\label{n_qubits_line}
	\mathbf{H}_{i}=\mathbf{n}_{i}^{(p)}\cdot \vec{\bm{\sigma}},
\end{align}
where $ p=1,2,3 $ and
\begin{align}
\mathbf{n}^{(p)}_{i}=\left\{
\begin{matrix}
&\mathfrak{R}\left( i\langle \psi_{1}|\bm{\sigma}_{p}|\psi_{2}\rangle  \right),\; i=1,\\
&\mathfrak{R}\left( i\langle \psi_{i}|\bm{\sigma}_{p}|\psi_{i+1}\rangle  \right)-\mathfrak{R}\left( i\langle \psi_{i-1}|\bm{\sigma}_{p}|\psi_{i}\rangle  \right),\\
& i=2,...,N-1,\\
&-\mathfrak{R}\left( i\langle \psi_{N-1}|\bm{\sigma}_{p}|\psi_{N}\rangle  \right),\; i=N.  \nonumber
\end{matrix}\right.
\end{align} 
Based on the quantum protocol (\ref{n_qubits_line}), we have the following result. 
\begin{Tem}\label{Th_Line_graph}
		Suppose that an $N$-qubit network is governed by (\ref{schr}), and Assumption 1 is satisfied. When using (\ref{n_qubits_line}),  the partial quantum consensus can be achieved globally. 
\end{Tem}
\begin{IEEEproof}
  Consider the Lyapunov function as follows
  \begin{align}\label{V}
V=&\frac{1}{2}\sum_{i=1}^{N-1}\langle \psi_{i}-\psi_{i+1} | \psi_{i}-\psi_{i+1} \rangle \nonumber\\
=&N-1-\frac{1}{2}\sum_{i=1}^{N-1}\mathfrak{R}(\langle \psi_{i}|\psi_{i+1} \rangle).
  \end{align}	
  By taking the derivative of $V$, we can obtain that 
  \begin{align}\label{dV}
  \dot{V}=&\sum_{i=1}^{N-1}-\mathfrak{R}\left( i\langle \psi_{i}|\mathbf{H}_{i}|\psi_{i+1}\rangle  \right)+\mathfrak{R}\left( i\langle \psi_{i}|\mathbf{H}_{i+1}|\psi_{i+1}\rangle  \right)\nonumber \\
  =&-\mathfrak{R}\left( i\langle \psi_{1}|\mathbf{H}_{1}|\psi_{2}\rangle  \right)\nonumber\\
  &+\sum_{i=2}^{N-1}\mathfrak{R}\left( i\langle \psi_{i-1}|\mathbf{H}_{i}|\psi_{i}\rangle  \right)-\mathfrak{R}\left( i\langle \psi_{i}|\mathbf{H}_{i}|\psi_{i+1}\rangle  \right)\nonumber\\
  &+\mathfrak{R}\left( i\langle \psi_{N-1}|\mathbf{H}_{N}|\psi_{N}\rangle  \right)\nonumber\\
  =&-\sum_{p=x,y,z}\mathbf{n}_{1}^{(p)}\cdot\mathfrak{R}\left( i\langle \psi_{1}|\bm{\sigma}_{p}|\psi_{2}\rangle  \right)\nonumber\\
  &+\sum_{i=2}^{N-1}\sum_{p=x,y,z}\mathbf{n}_{i}^{(p)}\cdot \Big(\mathfrak{R}\left( i\langle \psi_{i-1}|\bm{\sigma}_{p}|\psi_{i}\rangle  \right)\nonumber\\
  &-\mathfrak{R}\left( i\langle \psi_{i}|\bm{\sigma}_{p}|\psi_{i+1}\rangle  \right)  \Big)\nonumber\\
  &+\sum_{p=x,y,z}\mathbf{n}_{N}^{(p)}\cdot\mathfrak{R}\left( i\langle \psi_{N-1}|\bm{\sigma}_{p}|\psi_{N}\rangle  \right)\nonumber\\
  =&\;W_{1}+W_{2}+...+W_{N},
  \end{align}
  where $ W_{1}=-\sum_{p=x,y,z}\mathbf{n}_{1}^{(p)}\cdot\mathfrak{R}\left( i\langle \psi_{1}|\bm{\sigma}_{p}|\psi_{2}\rangle  \right) $, $ W_{i}=\sum_{p=x,y,z}\mathbf{n}_{i}^{(p)}\cdot \Big(\mathfrak{R}\left( i\langle \psi_{i-1}|\bm{\sigma}_{p}|\psi_{i}\rangle  \right)-\mathfrak{R}\left( i\langle \psi_{i}|\bm{\sigma}_{p}|\psi_{i+1}\rangle  \right)  \Big),\; i=2,...,N-1 $, $ W_{N}= \sum_{p=x,y,z}\mathbf{n}_{N}^{(p)}\cdot\mathfrak{R}\left( i\langle \psi_{N-1}|\bm{\sigma}_{p}|\psi_{N}\rangle  \right)$.

 Then, substituting (\ref{n_qubits_line}) into the above equation, we have $W_{i}\leq 0, \forall i=1,2,...,N$.
 Hence, the trajectories of the $N$-qubit network converge to the invariant set $S= \{|\psi \rangle \in \mathbb{C}^{2N} \; |\; W_{i}=0, i=1,2,...,N\}$.
 
  For the first term 
  \begin{align}
  W_{1}=&-\sum_{p=x,y,z}\Big(\mathfrak{R}\left( i\langle \psi_{1}|\bm{\sigma}_{p}|\psi_{2}\rangle  \right)\Big)^{2}\nonumber\\
  =&-\sin^{2}\frac{1}{2}\left(\theta_{1}-\theta_{2}\right)-\sin^{2}\frac{1}{2}\left(\phi_{1}-\phi_{2}\right)\cos^{2}\frac{1}{2}\left(\theta_{1}-\theta_{2}\right)\nonumber\\
  =&\;0, \nonumber
  \end{align}  
  we know that  $\theta_{1}=\theta_{2}+2k\pi$ and $\phi_{1}=\phi_{2}+2k\pi, k=0,1,2,...$, which means that  $|\psi_{1}\rangle = |\psi_{2}\rangle$. 
  Due to $\langle \psi_{i}|\bm{\sigma}_{p}|\psi_{i}\rangle =0$, $\forall \psi_{i}\in \mathbb{C}^{2}$, we can obtain that 
  \begin{align}
  W_2=-\sum_{p=x,y,z}\Big(\mathfrak{R}\left( i\langle \psi_{2}|\bm{\sigma}_{p}|\psi_{3}\rangle  \right)\Big)^{2}=0,  \nonumber
  \end{align}
  which follows that 
  $|\psi_{2}\rangle = |\psi_{3}\rangle$. 
  Repeating this procedure, one has 
  \begin{align}
  W_i=-\sum_{p=x,y,z}\Big(\mathfrak{R}\left( i\langle \psi_{i}|\bm{\sigma}_{p}|\psi_{i+1}\rangle  \right)\Big)^{2}=0, \forall i=3,...,N, \nonumber
  \end{align}
  which follows that $|\psi_{3}\rangle = |\psi_{4}\rangle=...=|\psi_{N}\rangle$. 
  Hence, we can conclude that the only equilibrium point in the invariant set $S$ is $|\psi_{1}\rangle = |\psi_{2}\rangle = ...=|\psi_{N}\rangle $, which means that the $|\psi_{i}(t)\rangle - |\psi_{j}(t)\rangle \rightarrow 0, i, j=1,...,N$ as time goes infinity. 
  
  		Next, we show that the Hamiltonian $\mathbf{H}_{i}=\mathbf{n}_{i}^{(p)}\cdot \bm{\sigma}$ designed above can be used in the local frame. 
  		Due to $|\psi_{i} \rangle =\mathbf{U}_{i} |\psi_{i} \rangle^{\mathcal{B}_{i}}$, we have $\frac{\text{d}}{\text{dt}}|\psi_{i}(t) \rangle^{\mathcal{B}_{i}}=-i\mathbf{H}_{i, \mathcal{B}_{i}}|\psi_{i}(t) \rangle^{\mathcal{B}_{i}}$, where $\mathbf{H}_{i, \mathcal{B}_{i}}=\mathbf{U}_{i}^{\dagger}\mathbf{H}_{i}\mathbf{U}_{i}$.
  	Recalling $\dot{V}$ in (\ref{dV}), we can consider $\langle \psi_{i} | \mathbf{H}_{i}| \psi_{j} \rangle$  and $\langle \psi_{i} | \mathbf{H}_{j}| \psi_{j} \rangle$ in the body frame.
  	
  	By substituting $\langle \psi_{i} |=^{\mathcal{B}_{i}}\langle \psi_{i} |\mathbf{U}_{i}^{\dagger}$, 
  		we have 
  		\begin{align}\label{relative_1}
  	\langle \psi_{i} | \mathbf{H}_{i}| \psi_{j} \rangle=&^{\mathcal{B}_{i}}\langle \psi_{i} |\mathbf{U}_{i}^{\dagger}\mathbf{U}_{i} \mathbf{H}_{i,\mathcal{B}_{i}} \mathbf{U}_{i}^{\dagger}\mathbf{U}_{j}| \psi_{j} \rangle^{\mathcal{B}_{j}}\nonumber\\
  	=&^{\mathcal{B}_{i}}\langle \psi_{i} | \mathbf{H}_{i,\mathcal{B}_{i}}  | \psi_{j} \rangle^{\mathcal{B}_{i}}, 	
  				\end{align}
  			where $ | \psi_{j} \rangle^{\mathcal{B}_{i}}=\mathbf{U}_{i}^{\dagger}\mathbf{U}_{j}| \psi_{j} \rangle^{\mathcal{B}_{j}}$ denotes the $| \psi_{j}\rangle$ expressed in the local frame $i$.
  			Similarly, we can derive that 
  			\begin{align}\label{relative_2}
  				\langle \psi_{i} | \mathbf{H}_{j}| \psi_{j} \rangle =&^{\mathcal{B}_{i}}\langle \psi_{i} |\mathbf{U}_{i}^{\dagger}\mathbf{U}_{j} \mathbf{H}_{j,\mathcal{B}_{j}} \mathbf{U}_{j}^{\dagger}\mathbf{U}_{j}| \psi_{j} \rangle^{\mathcal{B}_{j}}\nonumber\\
  				=&^{\mathcal{B}_{j}}\langle \psi_{i} | \mathbf{H}_{j,\mathcal{B}_{j}}  | \psi_{j} \rangle^{\mathcal{B}_{j}},
  				\end{align}
  			where $^{\mathcal{B}_{j}}\langle \psi_{i} |=^{\mathcal{B}_{i}}\langle \psi_{i} | \mathbf{U}_{i}^{\dagger}\mathbf{U}_{j}$ denotes the $| \psi_{i}\rangle$ expressed in the local frame $j$. 
  			From (\ref{relative_1}) and (\ref{relative_2}), the deriving procedure of the Hamiltonian $\mathbf{H}_{i}$ also works in the local frame of each qubit, which means that the Hamiltonian $\mathbf{H}_{i}$ can be used in the local frame $\mathcal{B}_{i}$ of each qubit.
\end{IEEEproof}

The above quantum protocol (\ref{n_qubits_line}) can also be used in an $2$-qubit system. 
For each qubit, the Hamiltonian is given as
\begin{align}\label{n_2_spins}
	\mathbf{H}_{i}=\mathbf{n}_{i}\cdot \vec{\bm{\sigma}},
\end{align}
where 
\begin{align}
\mathbf{n}_{i}= \left[
\begin{matrix}
\sin\frac{1}{2}(\phi_{i}+\phi_{j})\sin\frac{1}{2}(\theta_{i}-\theta_{j})\\
-\cos\frac{1}{2}(\phi_{i}+\phi_{j})\sin\frac{1}{2}(\theta_{i}-\theta_{j})\\
-\sin\frac{1}{2}(\phi_{i}-\phi_{j})\cos\frac{1}{2}(\theta_{i}-\theta_{j})
\end{matrix}\right], i,j=1,2, \nonumber
\end{align}
where $\theta_{i}\in[0 ,\pi]$ and $\phi_{i}\in [0, 2\pi)$.

In (\ref{n_2_spins}), the Hamiltonian is expressed in the world frame. 
We will also show that (\ref{n_2_spins}) can be also used in the local body frame, i.e., $\mathbf{H}_{i}^{\mathcal{B}}=\mathbf{n}_{i}^{\mathcal{B}}\cdot \vec{\bm{\sigma}}$.
{\color{black}The result is concluded in the following corollary.
\begin{Col}
			Suppose that a two-qubit system is governed by (\ref{schr}), when using the partial quantum protocol (\ref{n_2_spins}),  
			the partial quantum consensus can be achieved globally. 
\end{Col}}

{\color{black}\begin{Rmk}
		In Theorem 2, a Hamiltonian is designed for each qubit to achieve the quantum consensus based on Lyapunov-based approach. The physical realization of the protocol (\ref{n_qubits_line}) can be inspired from the experiments realizing quantum feedback through measurement-based feedback and coherent feedback in physical settings such as cavity QED \cite{2002_PRL}, opto-mechanics \cite{2017_pr_ZHANG}, superconducting circuits \cite{2013_PRL}, and quantum dots \cite{2010_PRL}.
		{\color{black}The Lyapunov-based approach has been used to design the quantum control feedback in \cite{Kuang2008}, and to achieve the asymptotic convergence. However, these works only consider the quantum state stabilization of a single quantum system. Different from these results, we design a quantum protocol for an $N$-qubit system. The main difficulty lies in the dynamics and the Hamiltonians of qubits being coupled.}
\end{Rmk}}
\begin{Rmk}
		In the above quantum consensus, we assume that the quantum state is a product state, and the weak measurement is performed on each subsystem. 
		For the general entangled state, the measurement performed on the state of each subsystem will affect the state of other nodes. In fact, there are some existing results focusing on this problem \cite{Jacobs2006,2007_Optimal}.
		In \cite{2010_TAC_J.Zhang}, a quantum feedback control is designed based on the partial information extracted from quantum weak measurements. 
		Through the quantum feedback control, the coherence and entanglement of quantum systems can be partially preserved.
		Inspired by this technique, we can exploit the quantum consensus protocol while protecting the coherence and entanglement of $N$-qubit systems.
		The detailed procedure is shown in Appendix A.
		However, the result in \cite{2010_TAC_J.Zhang} only considers the single qubit system and the two-qubit system. In addition, the Hamiltonian design for two-qubit systems is non-local. How to design the local Hamiltonian for $N$-qubit systems to preserve the entanglement is still an open problem to be solved. 
	\end{Rmk}
	\begin{Rmk}
	{\color{black}	Recently, the data-driven robust quantum control has been investigated in \cite{2020_TCYB_Dong,2017_Quantum_Dong_TCYB,2019_TCYB_donglearning}. 
		A robust quantum unitary transformation is computed based on a gradient flow algorithm, and
		an optimal control strategy is find by a training process based on the data sampled previously \cite{2017_Quantum_Dong_TCYB}.
		Compared with the data-driven quantum control method in \cite{2017_Quantum_Dong_TCYB}, the unitary transformation (\ref{n_qubits_line}) is designed by using local quantum states obtained by weak measurements without using other additional information.}
	\end{Rmk}
{\color{black}\subsection{Geometry-based quantum protocol design}}
In this part, we consider the more general graph case where the topology is a connected graph.
There are some situations where the particles are trapped in a lattice or a two-dimensional space.

{\color{black}In the connected graph case, it is difficult to derive the Hamiltonian to guarantee that the invariant set $S$ consists of only the consensus point, i.e., $|\psi_{1}\rangle = |\psi_{2}\rangle = ...=|\psi_{N}\rangle $ using the Lyapunov function in (\ref{V}).}
Recalling that in Theorem 1, we have derived the rotation axis $\mathbf{n}_{12}$ of two qubits by using the cross product of two direction vectors in three-dimensional spherical coordinates. 
The two direction vectors correspond to the two quantum states on the Bloch ball. 
This fact inspires us to investigate the quantum consensus protocol from a geometric point of view.

For an agent whose state is a three-dimensional vector $\mathbf{x}_{i}\in \mathbb{R}^{3}$ evolving on the 2-sphere, we obtain its dynamics as follows, 
\begin{align}{\label{dynamic_2-sphere}}
	\dot{\mathbf{x}}_{i}=(\mathbf{I}_{3}-\mathbf{x}_{i}\mathbf{x}_{i}^{\top})\mathbf{u}_{i}, 
\end{align}
where $\mathbf{u}_{i}$ is the velocity input and $(\mathbf{I}_{3}-\mathbf{x}_{i}\mathbf{x}_{i}^{\top})$ is a projection which projects the input $\mathbf{u}_{i}$ to the tangent space of the 2-sphere located at $\mathbf{x}_{i}$.
To achieve the consensus on the 2-sphere, i.e., $\mathbf{x}_{1}=...=\mathbf{x}_{N}$, we can construct the consensus protocol of each agent as follows,
\begin{align}\label{sphere_consensus}
	\mathbf{u}_{i}=\sum_{j\in \mathcal{N}_{i}(t)}w_{ij}\mathbf{x}_{j},
\end{align}
where $w_{ij}(\cdot)$ is the weighted function depending on the Euclidean distance $\|\mathbf{x}_{i}-\mathbf{x}_{j}\|$.
The protocol (\ref{sphere_consensus}) can be considered as a weighted sum of neighbors' states which is projected onto the tangent space of 2-sphere located at $\mathbf{x}_{i}$ \cite{Markdahl2018}. 

Next, we build a connection between the partial quantum consensus to the consensus on the 2-sphere.

\begin{Lma}
	Suppose the qubit dynamics governed in (\ref{schr}). 
	Let $\mathbf{x}_{i}$ be a vector in the three-dimensional sphere. 
	The dynamic equation (\ref{schr}) can be transformed into the following form
	\begin{align}\label{dx}
	  \dot{\mathbf{x}}_{i}=\left(\mathbf{n}_{i}\cdot\vec{\mathbf{J}}\right)\mathbf{x}_{i},
	\end{align}
	where $\mathbf{n}_{i}\cdot\vec{\mathbf{J}}=n_i^{(1)}\mathbf{J}_{1}+n_i^{(2)}\mathbf{J}_{2}+n_i^{(3)}\mathbf{J}_{3}$, and 
	\begin{align}
	\mathbf{J}_{1}=\left[
	\begin{matrix}
	0&0&0\\
	0&0&-1\\
	0&1&0\\
	\end{matrix}\right],
		\mathbf{J}_{2}=\left[
	\begin{matrix}
	0&0&1\\
	0&0&0\\
	-1&0&0\\
	\end{matrix}\right],
		\mathbf{J}_{3}=\left[
	\begin{matrix}
	0&-1&0\\
	1&0&0\\
	0&0&0\\
	\end{matrix}\right]. \nonumber
	\end{align}
\end{Lma}
\begin{IEEEproof}
	In (\ref{spin}), we know that $|\psi_{i}\rangle\in \mathbb{C}^{2}$ is a 2-dimensional complex vector such that $\langle\psi_{i}|\psi_{i}\rangle=1$.  
	According to the dynamics (\ref{schr}), the evolution of $|\psi_{i} (t)\rangle$ is a unitary transformation
	\begin{align}
	|\psi_{i}(t)\rangle = \mathbf{U}_{i}(t)|\psi_{i}(0)\rangle,
	\end{align}
	where $\mathbf{U}_{i}(t)\in \mathbb{SU}(2)$ is a unitary matrix.  
	Alternatively, we can equivalently express the element in $\mathbb{S}^{2}$ in Cartesian coordinate, i.e., $\mathbf{x}_{i}\in \mathbb{R}^{3}$ and $ \|\mathbf{x}_{i}\|_{2}=1$. 
	In Cartesian coordinate, the rotation is denoted as a three-dimensional orthogonal matrix $\mathbf{R}_{i}\in \mathbb{SO}(3)$, i.e., 
		\begin{align}\label{x(t)}
	\mathbf{x}_{i}(t) = &\;\mathbf{R}_{i}(t)\mathbf{x}_{i}(0)\nonumber\\
	=&\;e^{\mathbf{n}_{i}\cdot \vec{\mathbf{J}}t}\mathbf{x}_{i}(0),
	\end{align}
	where $\mathbf{n}_{i}\cdot \vec{\mathbf{J}}$ is the skew symmetric matrix expanded by the basis $\vec{\mathbf{J}}$ in  $\mathfrak{so}(3)$, and $\mathbf{n}_{i}\in \mathbb{R}^{3}$ is the unit vector which represents the rotation axis.	
	From (\ref{x(t)}), we have 
$
	   \dot{\mathbf{x}}_{i}(t)=\left(\mathbf{n}_{i}\cdot \vec{\mathbf{J}}\right)\mathbf{x}_{i}(t),
$ which completes the proof.
	\end{IEEEproof}

Based on Lemma 1 and there is a one-to-one mapping from $|\psi_{i}\rangle \in \mathbb{C}^{2}$ to $\mathbf{x}_{i}\in \mathbb{S}^{2}$, 
we can actually construct the rotation $\mathbf{R}_{i}$ by designing the rotation axis $\mathbf{n}_{i}$ in (\ref{dx}), which is equivalent to determining the Hamiltonian in (\ref{schr}).
Inspired by the protocol (\ref{sphere_consensus}), we can construct the quantum consensus protocol below,
\begin{align}\label{n_n_spins_connected}
	\mathbf{H}_{i}=	\mathbf{n}_{i}\cdot \vec{\bm{\sigma}},
\end{align}
where 
$
	\mathbf{n}_{i}= \;\mathbf{x}_{i}\times (\mathbf{I}_{3}-\mathbf{x}_{i}\mathbf{x}_{i}^{\top})\sum_{j\in \mathcal{N}_{i}}w_{ij}\mathbf{x}_{j}. \nonumber
$
Now, we can state the main theorem in the following.
\begin{Tem}\label{Th_switching_connected}
	Suppose that an $N$-qubit network  is described by (\ref{schr}) under a connected graph. When using (\ref{n_n_spins_connected}), if the initial quantum states are contained in the hemisphere on the Bloch ball, then 
	the partial quantum consensus is achieved.  In addition, the quantum protocol (\ref{n_n_spins_connected}) can be also used in the local frame of each qubit, i.e., $\mathbf{H}_{i}^{\mathcal{B}}=\mathbf{n}_{i}^{\mathcal{B}}\cdot \vec{\bm{\sigma}}$. 
\end{Tem}

\begin{IEEEproof}
	We will establish the relationship between the dynamics (\ref{dx}) and (\ref{sphere_consensus}). 
	 Since $\left(\mathbf{n}_{i}\cdot\vec{\mathbf{J}}\right)$ is a real three-dimensional skew symmetric matrix, we have $\left(\mathbf{n}_{i}\cdot\vec{\mathbf{J}}\right)\mathbf{x}_{i}=\mathbf{n}_{i}\times \mathbf{x}_{i}$. 
	 Recalling (\ref{dx}), one has 
	 \begin{align}\label{n}
	  \dot{\mathbf{x}}_{i}=\left(\mathbf{n}_{i}\cdot \vec{\mathbf{J}}\right)\mathbf{x}_{i}= \mathbf{n}_{i}\times \mathbf{x}_{i}.
	 \end{align} 
	  From (\ref{n}), we know that the vectors $\mathbf{n}_{i}$, $\mathbf{x}_{i}$, and $\dot{\mathbf{x}}_{i}$ are mutually orthogonal. 
	  By using (\ref{n_n_spins_connected}) and based on the fact that each qubit state $|\psi_{i}\rangle$ corresponds to a state $\mathbf{x}_{i}(\theta_{i},\phi_{i})$, 
	  we can obtain that the dynamics of $\mathbf{x}_{i}$ in (\ref{dx}) can be transformed into (\ref{dynamic_2-sphere}).
	  Then, inspired by the results in \cite{Markdahl2018}, we can conclude that the hemisphere is invariant and 
	  the only stable equilibrium point of the system (\ref{sphere_consensus}) is the consensus point in  $\{\mathbf{x}\in \mathbb{R}^{3N}\;| \;\mathbf{x}_{i}=\mathbf{x}_{j}, i, j=1,...,N\}$. Hence, 
	  the asymptomatic consensus is achieved under connected graphs such that $\mathbf{x}_{i}(t)\rightarrow\mathbf{x}_{j}(t), i, j=1,...,N$ as $t\rightarrow +\infty$ for all $\mathbf{x}_{i}(0)$ contained in the hemisphere. Due to the one-to-one corresponding of $\mathbf{x}_{i}(t)$ and $|\psi_{i}(t)\rangle$ in the Bloch ball, we know that the 
	  $|\psi_{i}(t)\rangle \rightarrow |\psi_{j}(t)\rangle, i, j=1,...,N$ as $t 
	  \rightarrow +\infty$ for all $|\psi_{i}(0)\rangle$ contained in the hemisphere.
	  \begin{figure}[htb]	
	  	\centering	
	  	{
	  		\includegraphics[scale=0.6]{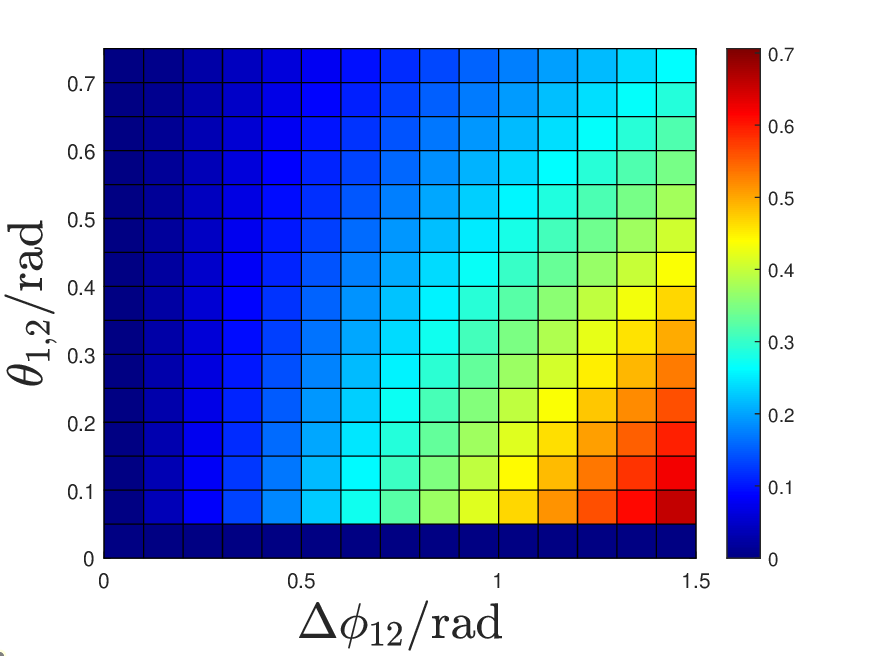}		
	  	}
	  	\caption{The difference value $T_{1}-T_{\min}$ between the settling time of the protocol and the minimum time on the Bloch ball. The $y$-axis and $x$-axis are defined as  $\theta_{1,2}(0)=\theta_{1}(0)=\theta_{2}(0)$ and $\Delta \phi_{12}(0)=\phi_{2}(0)-\phi_{1}(0)$ where the initial values of polar angles and azimuthal angles are set as $\theta_{1}(0)=\theta_{2}(0)\in (0,\frac{\pi}{4})$ and $\phi_{1}(0), \phi_{2}(0)\in(0,\frac{\pi}{2})$. The settling time $T_{1}$  is chosen as $T_{1}=\inf_{t>0}\{t\in \mathbb{R} : V(t)<10^{-5}\}$ and the minimum time $T_{\min}$ is calculated by ($\ref{Tmin}$).
	  	}
	  	\label{delta_T}
	  \end{figure}
	  \begin{figure*}[htb]	
	  	\centering
	  	{
	  		\includegraphics[scale=0.4]{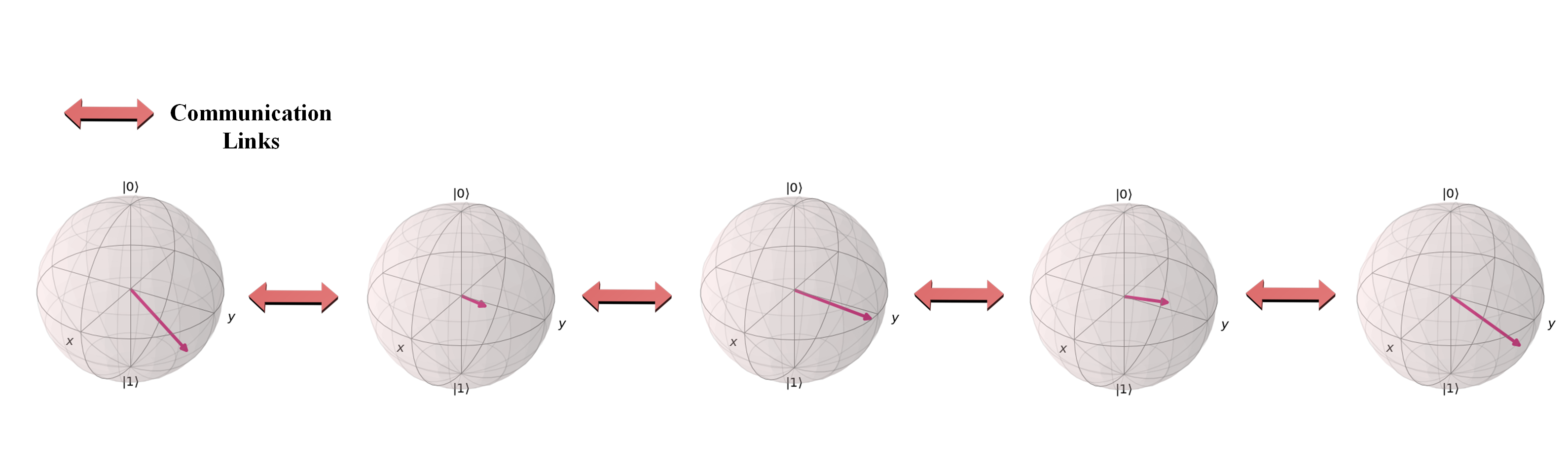}	
	  	}
	  	\caption{{\color{black}A qubit network} with a chain connected topology. }	\label{line_graph}
	  \end{figure*}
  
	{\color{black}  Next, we show that the designing rotation axis $\mathbf{n}_{i}$ in (\ref{n_n_spins_connected}) can be used in the local body frame $ \mathcal{B}_{i}$, i.e.,
	  \begin{align}{\label{relative_n}}
	  \mathbf{n}_{i}^{\mathcal{B}_{i}}=\;& \mathbf{x}_{i}^{\mathcal{B}_{i}} \times (\mathbf{I}_{3}-\mathbf{x}_{i}^{\mathcal{B}_{i}}{\mathbf{x}_{i}^{\mathcal{B}_{i}}}^{\top})\sum_{j\in \mathcal{N}_{i}}w_{ij}\mathbf{x}_{j}^{\mathcal{B}_{i}},
	  \end{align}
  where $\mathbf{x}_{i}^{\mathcal{B}_{i}}=\mathbf{R}_{i}^{\top}\mathbf{x}_{i}^{\mathcal{W}}$ and $\mathbf{x}_{j}^{\mathcal{B}_{i}}$ is the qubit $j$'s state measured in the qubit $i$'s local frame $\mathcal{B}_{i}$. }
	  Recalling the dynamics (\ref{dx}) and $\mathbf{R}_{i}\mathbf{x}_{i}^{\mathcal{B}_{i}}=\mathbf{x}_{i}^{\mathcal{W}}$, where $ \mathbf{R}_{i} $ is the rotation from the world frame $ \mathcal{W}$ to the body frame $ \mathcal{B}_{i}$, one has 
	  \begin{align}{\label{relative_n_1}}
	    \dot{\mathbf{x}}_{i}^{\mathcal{B}_{i}}=&\left( n_{i}^{(1)}\mathbf{R}_{i}^{\top}\mathbf{J}_{1}\mathbf{R}_{i}+ n_{i}^{(2)}\mathbf{R}_{i}^{\top}\mathbf{J}_{2}\mathbf{R}_{i}+n_{i}^{(3)}\mathbf{R}_{i}^{\top}\mathbf{J}_{3}\mathbf{R}_{i}\right)\mathbf{x}_{i}^{\mathcal{B}_{i}}\nonumber\\
	    =&\left(\sum_{k=1}^{3}n_{i}^{(k)}\text{Skew}(\mathbf{R}_{i}^{\top}\mathbf{e}_{i}^{\mathcal{W}})\right)\mathbf{x}_{i}^{\mathcal{B}_{i}}\nonumber\\
	    =&\left(\sum_{k=1}^{3}n_{i}^{(k)}\mathbf{J}_{k}^{\mathcal{B}_{i}}\right)\mathbf{x}_{i}^{\mathcal{B}_{i}}\nonumber\\
	    =&\;\mathbf{n}_{i}^{\mathcal{B}_{i}}\times \mathbf{x}_{i}^{\mathcal{B}_{i}},
	  \end{align}
	  where we use the facts that  $\mathbf{R}_{i}^{\top}\mathbf{e}_{1}^{\mathcal{W}}=\mathbf{e}_{1}^{\mathcal{B}_{1}}$ to obtain the third equality. The $ \mathbf{e}_{1}^{\mathcal{B}_{1}}$ is the first column vector of the matrix $ \mathbf{I}_{3} $.  
	  For (\ref{sphere_consensus}), we can also derive that 
	  \begin{align}{\label{relative_n_2}}
	  \dot{\mathbf{x}}_{i}^{\mathcal{B}_{i}}=&\;\mathbf{R}_{i}^{\top}\dot{\mathbf{x}}_{i}\nonumber\\
	  =&\;\mathbf{R}_{i}^{\top}\left(\mathbf{I}_{3}-\mathbf{R}_{i}\mathbf{x}_{i}^{\mathcal{B}_{i}}{\mathbf{x}_{i}^{\mathcal{B}_{i}}}^{\top}\mathbf{R}_{i}^{\top}\right)\mathbf{R}_{i}\sum_{j\in \mathcal{N}_{i}}w_{ij}\mathbf{R}_{i}^{\top}\mathbf{R}_{j}\mathbf{x}_{j}^{\mathcal{B}_{j}}\nonumber\\
	  =&\;(\mathbf{I}_{3}-\mathbf{x}_{i}^{\mathcal{B}_{i}}{\mathbf{x}_{i}^{\mathcal{B}_{i}}}^{\top})\sum_{j\in \mathcal{N}_{i}}w_{ij}\mathbf{x}_{j}^{\mathcal{B}_{i}}.
	  \end{align}
	  Hence, by combing (\ref{relative_n_1}) and (\ref{relative_n_2}), we can derive ({\ref{relative_n}}), which 
	  shows that the consensus protocol (\ref{n_n_spins_connected}) is not dependent on the coordinate frame. 
	  The proof is completed.
\end{IEEEproof}

%
\begin{Rmk}
		{\color{black}Most of the existing research on quantum consensus \cite{Ticozzi2016,Shi2016,Shi2017,Jafarizadeh2016,Jafarizadeh2017} describes the interaction between qubits by the swapping operator, and the symmetric quantum consensus and reduced state quantum consensus can be achieved. However, the swapping operator is a quantum global operation which cannot be implemented in a distributed manner. 
    Thus, we design the Hamiltonian for each quantum system only depending on the local information by using the Lyapunov method and geometry method, respectively.
	In addition, we also show that the proposed Hamiltonian for each quantum can be used in its own body frame, which removes the requirement of a global frame for partial quantum consensus of qubit networks}. 
\end{Rmk}
	\begin{Rmk}
	{\color{black}	In this section, we propose two partial quantum consensus protocols based on the Lyapunov method and the geometry method, respectively. Both of them have its benefits. 
	The quantum protocol designed by the geometry method can be used in more general graphs. However, the initial quantum states of qubit networks should be contained in the hemi-sphere due to the geometry constraint on $\mathbb{S}^{2}$, which implies that the consensus result is local.
	For a particular chain connected graph, the consensus protocol can be designed by the Lyapunov method and the global partial quantum consensus can be achieved if initial quantum states are contained in the Bloch ball.
		It should be noted that achieving partial quantum consensus globally with general connected graphs is an open and challenging problem to be studied further. }
	\end{Rmk}

 \section{Simulation}

  In this section, we are firstly interested in evaluating the convergent settling time of the Hamiltonian (\ref{n_2_spins}) compared with the minimum time estimation in Theorem 1. 
  Given a two-qubit system, we set the initial quantum states of both qubits as $\theta_{1}(0)=\theta_{2}(0)\in(0,\frac{\pi}{4})$, $\phi_{1}(0), \phi_{2}(0)\in(0,\frac{\pi}{2})$ and define
  $\Delta \phi_{12}=\phi_{2}(0)-\phi_{1}(0)$.
  By using the Hamiltonian (\ref{n_2_spins}), the function $V=\frac{1}{2}
  \langle \bm{\psi}_{1}-\bm{\psi}_{2} | \bm{\psi}_{1}-\bm{\psi}_{2} \rangle$ will tend to zero as $t \rightarrow \infty$. 
  In order to make the comparison easy, we choose the settling time $T_{1}=\inf_{t>0}\{t\in \mathbb{R} : V(t)<10^{-5}\}$ in the numerical simulation. 
In Fig. \ref{delta_T},  it shows the difference value of the settling time of the Hamiltonian (\ref{n_2_spins})  compared with the minimum time estimation calculated by  $T_{1}-T_{\min}$.
We observe that the settling time of the protocol (\ref{n_2_spins}) 
is very close to the minimum time in most regions on the Bloch ball, especially 
when the initial values of $\theta_{1}$ and $\theta_{2}$ are close to $\frac{\pi}{2}$ (the points located at the equator of the Bloch ball).
The only exception is when $|\bm{\psi}_{1}\rangle(0)$ and $|\bm{\psi}_{2}\rangle(0)$ are located near the north pole. In this case, we find that the settling time of the protocol (\ref{n_2_spins}) is larger than the minimum time. This is because the protocol (\ref{n_2_spins}) is designed based on the Euclidean distance between two quantum states which is not always the shortest distance on the Bloch ball.
  \begin{figure}[htb]	
  	\centering	
  	{
  		\includegraphics[scale=0.20]{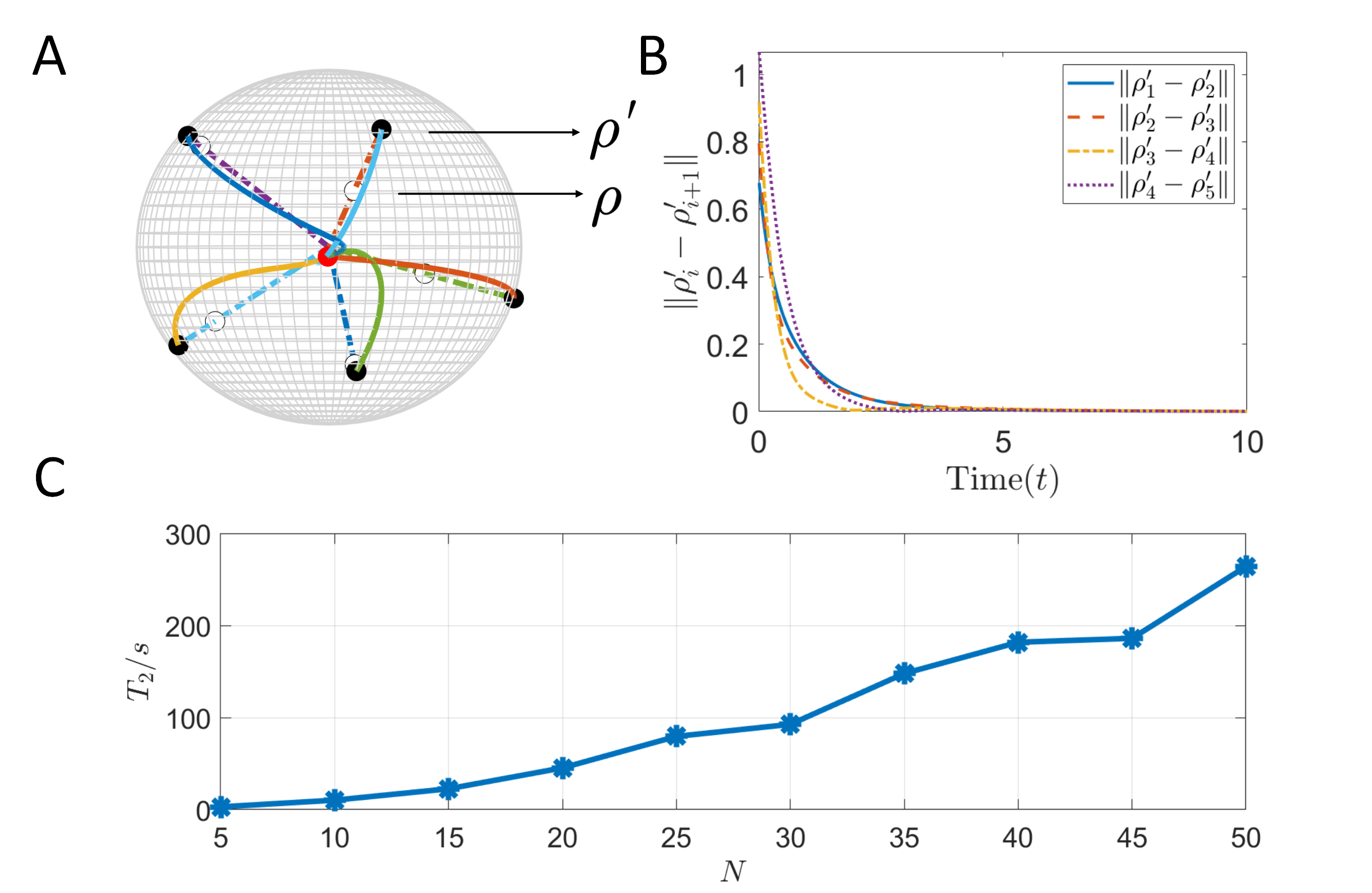}
  	}
  	\caption{The partial quantum consensus result under chain graphs. A) The trajectories of $5$ qubits under chain graphs evolving on the Bloch ball.  
  		B) The norm of error for the density matrix of quantum states under chain graphs. 
  		C) The settling time $T_{2}$ of partial quantum consensus for qubit networks with the different number of qubits.}
  	\label{qubits_Line}
  \end{figure}

\begin{figure}[b]
	\centering
	{
		\includegraphics[scale=0.25]{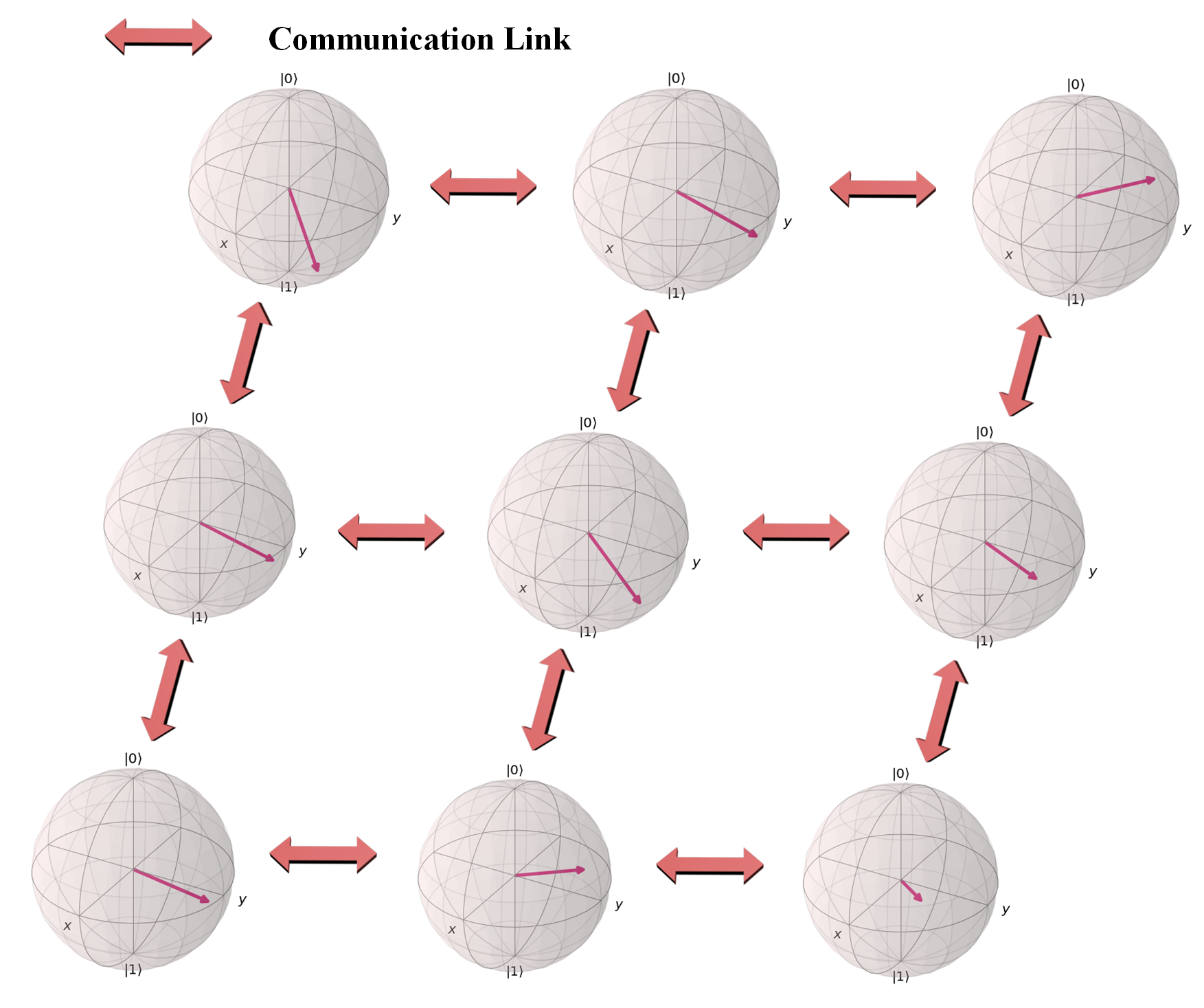}		
	}
	\caption{A 9-qubit network with connected topologies. }	\label{9_qubits}
\end{figure}
  {\color{black}Then, we simulate a five-qubit network with a chain graph. The initial states located on the Bloch ball are shown in Fig. \ref{line_graph}. 
  The white ball denotes the initial mixed quantum state and the black ball on the surface of the Bloch ball denotes its part of the pure state. }
  Then, by using the quantum protocol (\ref{n_qubits_line}), the trajectories of qubits evolving on the Bloch ball are shown in Fig. \ref{qubits_Line}A and the norm of error for the density matrix of the pure state is shown in Fig. \ref{qubits_Line}B. 
  It can be seen that all quantum states are located in the same direction on the Bloch ball, which implies that the partial quantum consensus is achieved. 
In Fig. \ref{qubits_Line}C, it shows the settling time of the partial quantum consensus for qubit networks with the different number of qubits.
 Let $V(t)=\frac{1}{2}\sum_{i=1}^{N-1}\langle \bm{\psi}_{i}-\bm{\psi}_{i+1} | \bm{\psi}_{i}-\bm{\psi}_{i+1} \rangle$.
  The settling time $T_{2}$ is defined as 
  $T_{2}=\inf_{t>0}\{t\in \mathbb{R} : V(t)<10^{-2}\}$.
  We can easily observe that the settling time will not exponentially increase when the scale of networks becomes larger. 
  
\begin{figure}[htb]	
	\centering	
	{
		\includegraphics[scale=0.20]{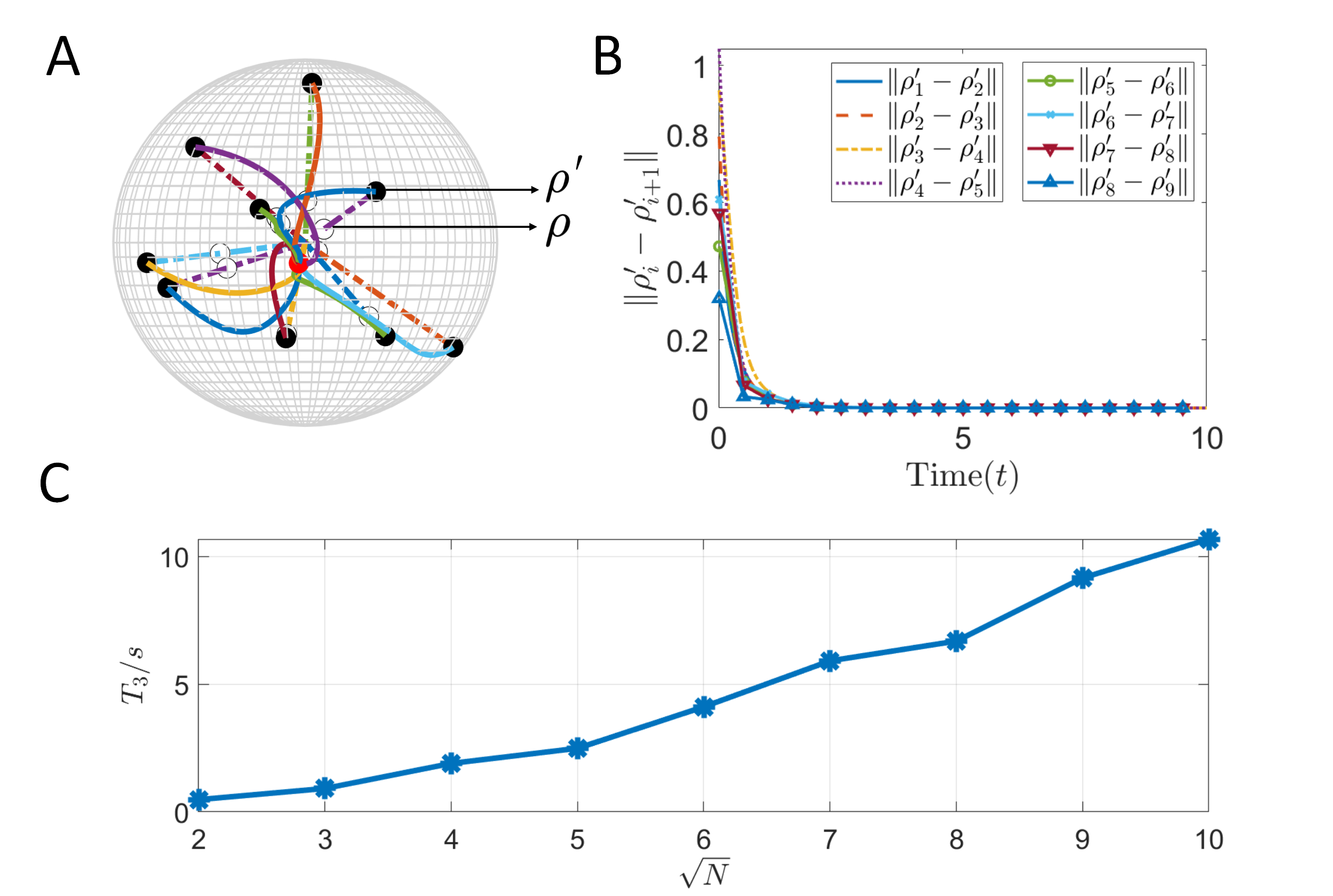}
	}
	\caption{The partial quantum consensus result under connected graphs. A) The trajectories of $9$ qubits under connected graphs evolving on the Bloch ball. 
		B) The norm of error for the density matrix of quantum states under connected topologies.  
		C) The settling time $T_{3}$ of partial quantum consensus for qubit networks with the different number of qubits.}
	\label{qubits_connected}
\end{figure}
   
  Next, we consider an example of a 9-qubit network with a connected graph. 
  The topology graph is a $3\times3$ grid shown in Fig. \ref{9_qubits}. 
  The initial quantum states are illustrated by the red 
  arrow on the Bloch ball. 
  By using the protocol (\ref{n_n_spins_connected}), the trajectories of each qubit evolving on the Bloch ball are shown in Fig. \ref{qubits_connected}A
  and the norm of error for the density matrix of the pure state is shown in Fig. \ref{qubits_connected}B. 
  It is clear that the partial quantum consensus is achieved.

  In Fig. \ref{qubits_connected}C, it shows the relationship between the settling time of the partial quantum consensus and the scales of qubit networks.
  The settling time $T_{3}$ is defined as 
  $T_{3}=\inf_{t>0}\{t\in \mathbb{R} : V(t)<10^{-2}\}$.
  Without losing generality, we assume that the topology graph is a grid with $n \times n$ nodes, where $n=2,3,4,...,10$ in the simulation. 
  From Fig. \ref{qubits_connected}E, we know that the settling time increases when the scale of networks becomes larger. 
  In addition, by comparing the results in Fig. \ref{qubits_Line}E, we can find that the settling time of partial quantum consensus for connected graphs is much smaller than chain graphs. 
  {\color{black}This is reasonable since the algebraic connectivity of a grid graph considered in the simulation is larger than the chain connected graph. 
  	Note that the algebraic connectivity of network topology, i.e., the second smallest eigenvalue of Laplacian, is a measure of convergence speed of consensus algorithms.}

    {\color{black}We further compared the convergence performance of the proposed quantum consensus protocols with QCMEs in \cite{Shi2016,Shi2017,Jafarizadeh2017}. 
    Here, we consider a 3-qubit network with a chain topology and a connected graph, respectively, to facilitate the comparison of three kinds of consensus protocols. 
    The composite state of the multi-agent system is defined as $\bm{\rho}=\bm{\rho}^{\prime}_{1}\otimes\bm{\rho}^{\prime}_{2}\otimes\bm{\rho}^{\prime}_{3} \in \mathbb{C}^{8 \times 8}$, where $\bm{\rho}^{\prime}_{i},\; i=1,2,3$ is the pure state of each qubit. Let the quantum average of the	3-qubit network be $\bar{\bm{\rho}}$.  
    For the chain graph, the edge set is given as
    $\mathcal{E}_{l}=\{(1,2),(2,1),(2,3),(3,2)\}$.
    We will compare the convergence rate of three different protocols in terms of the 2-norm of the matrix $(\bm{\rho}-\bar{\bm{\rho}})$.
    The trajectory of the value of $\|\bm{\rho}-\bar{\bm{\rho}}\|$ is shown in Fig. 7A. 
    It can be seen that the protocol  (\ref{n_n_spins_connected}) converges to the consensus state faster than QCMEs while the protocol (\ref{n_qubits_line}) converges slower than QCMEs in Fig. 7A. 
    For the connected graph, the edge set is given as $\mathcal{E}_{c}=\{(1,2),(2,1),(2,3),(3,2),(3,1),(1,3)\}$.
    We will compare the convergence rate of QCMEs and the protocol  (\ref{n_n_spins_connected}). 
   {\color{black} Figure 7B shows that the protocol (\ref{n_n_spins_connected}) converge faster than QCMEs. 
    It should be noted that we only compare the converge rate numerically here. The rigorous comparison with the theoretical analysis is worth studying in the future work.}
}
\begin{figure}[htb]		
		\centering	
		{
				\includegraphics[scale=0.13]{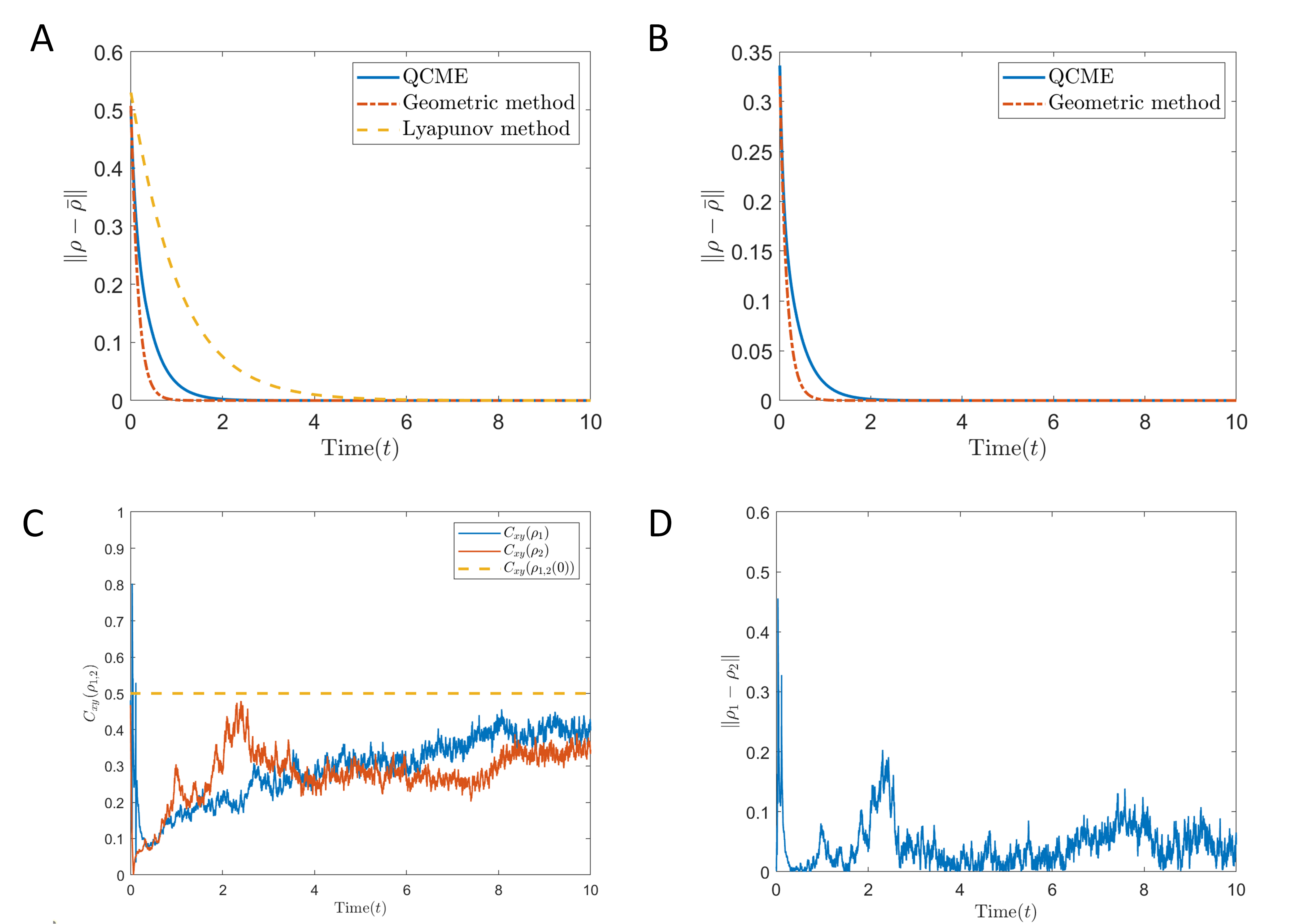}
				\label{convergence_rate_1}
			}
		\caption{A) Comparisons of the convergence rate under chain graphs. B) Comparisons of the convergence rate under connected graphs. C) Plot of coherence function $C_{xy}(\bm{\rho}_{1,2})$. D) Distance between  $\bm{\rho}_{1}$ and $\bm{\rho}_{2}$.}
	\end{figure}

\section{Conclusion}
In this paper, a partial quantum consensus problem is considered for qubit networks. 
Based on the convexity property of the density matrix, we construct the unitary transformation for each quantum system to drive the part of pure states to consensus. 
The two-qubit system is firstly studied by using the geometry method. 
{\color{black}The minimum completing time for the unitary transformation is derived based on the geometric configuration of the quantum state.} 
To obtain an asymptomatic consensus result, we design the Hamiltonian for each quantum system based on the Lyapunov approach. 
Then, we extend the approaches to the more general $N$-qubit network. 
For a chain topology case, the Hamiltonian of each quantum system is designed based on the Lyapunov approach.
For the general connected graph case, we exploit a geometric method which builds the relationship between partial quantum consensus and 2-sphere consensus problem. 
Finally, a numerical simulation for 2-qubit and 9-qubit quantum networks is demonstrated to show the effectiveness of theoretical results.

{\color{black}Future works may concentrate on three aspects. 
	Firstly, this work focuses on qubit networks in two-dimensional Hilbert space $\mathcal{H}_{2}$, the more general problem is the quantum consensus on the Hilbert space with higher dimensions such as $\mathcal{H}_{2^{n}}, \;n > 1$.
	Secondly, the complete quantum consensus for qubit networks should be further investigated. The main idea is to find a local Lindblad operator for each agent which only depends on the information of the quantum state of itself and its neighbors.   
	Thirdly, 
	the multi-agent deep reinforcement learning has shown a great potential in the multi-agent framework where multiple agents learn to act optimally in a shared environment. 
	A typical limitation of the multi-agent deep reinforcement learning (MADRL) algorithm is the long training time and high requirement of the computation resource. Thus, the quantum multi-agent deep reinforcement learning (QMADRL) is a promising direction which offers a novel method to deal with this problem owing to its strong ability of parallel computing.}


%

\appendices

{\color{black}\section{Feedback control protocol design for coherence protection of two-qubit systems}
%
    Due to the disturbance induced by weak measurements, we consider the system dynamics represented by the following master equation \cite{2010_TAC_J.Zhang},
\begin{align}\label{Master}
	d \bm{\rho}_{i}=&-i\left[\mathbf{H}_{i}, \bm{\rho}_{i}\right] d t+4 \Gamma_{r} \mathcal{D}\left[\bm{\sigma}_{-}\right] \bm{\rho}_{i} dt\nonumber\\
	&+\left(\Gamma_{\phi}+\Gamma_{z}\right) \mathcal{D}\left[\bm{\sigma}_{z}\right] \bm{\rho}_{i} d t+\sqrt{\eta_{z} \Gamma_{z}} \mathcal{H}\left[\bm{\sigma}_{z}\right] \bm{\rho}_{i} dW, \nonumber\\
	dy_{z}^{i}=&\left\langle\bm{\sigma}_{z}\right\rangle_{\bm{\rho}_{i}} d t+\frac{1}{2 \sqrt{\eta_{z} \Gamma_{z}}} d W,\; i=1,2.
\end{align}
In (\ref{Master}), the last three terms represent the influence of environmental disturbances including relaxation and dephasing effects. 
The Hamiltonian $\mathbf{H}_{i}$ aims to preserve the value of the coherence function $C_{xy}(\bm{\rho}_{i})$ defined as 
$
	C_{x y}(\bm{\rho}_{i})=\sqrt{\left\langle\bm{\sigma}_{x}\right\rangle_{\bm{\rho}_{i}}^{2}+\left\langle \bm{\sigma}_{y}\right\rangle_{\bm{\rho}_{i}}^{2}}, \;i=1,2.
$
The second equation in (\ref{Master}) is the measurement outcome. 
Next, by using information obtained from weak measurements and the initial state, we design the Hamiltonian $\mathbf{H}_{i}$ for two-qubit systems to protect the coherence.
The Hamiltonian adjusted by the time-varying control $\mu_{i}$ for two-qubit systems can be designed as $\mathbf{H}_{i}=\mu_{i}\left(\cos \phi_{ij} \bm{\sigma}_{x}+\sin \phi_{ij} \bm{\sigma}_{y}\right)$, $\mu_{i}=\frac{\Gamma C_{xy}(\bm{\rho}_{i}(0))}{Y_{z}^{i}}$, 
$\phi_{ij}=0.5\arctan \left(-\frac{\left\langle\bm{\sigma}_{x}\right\rangle_{\rho_{i}}}{\left\langle\bm{\sigma}_{y}\right\rangle_{\rho_{i}}}\right)+0.5\arctan \left(-\frac{\left\langle\bm{\sigma}_{x}\right\rangle_{\rho_{j}}}{\left\langle\bm{\sigma}_{y}\right\rangle_{\rho_{j}}}\right)$, and $Y_{z}^{i}=\frac{1}{t} \int_{0}^{t}\left\langle\bm{\sigma}_{z}(\tau)\right\rangle_{\bm{\rho}_{i}} d \tau+\frac{1}{2 \sqrt{\eta_{z} \Gamma_{z}} t} \int_{0}^{t} d W,
\;	i,j=1, 2, i\neq j.$

To show the effectiveness of the control strategy, 
we choose $\Gamma_{r}=\Gamma_{\phi}=10$, $\Gamma_{z}=0.1$,  $\Gamma=\Gamma_{r}+\Gamma_{\phi}+\Gamma_{z}$, and $\eta_{z}=1$.
Under the feedback control, the numerical result is shown in Figs. 7C-7D.
In Fig. 7C, the solid lines are trajectories of coherence functions of two qubits under feedback control and the dash line is the target trajectories without feedback control and decoherence.
It shows that coherence values of two qubits tend to the target trajectory of the initial coherence. This result shows that the coherence of two qubits are protected effectively. {\color{black}In Fig. 7D, 
the Euclidean distance between $\rho_1$ and
$\rho_2$ is decreasing as time goes to infinity. Due to the influence of environmental noises, the Euclidean distance may not converge to zero precisely. Note that the consensus error is dependent on the mean value of environmental noises.}
 $\hfill\blacksquare$}
%
%
%
\section*{Acknowledgment}
The authors would like to thank Wangli He and Zhixing Cao for helpful discussions.

\bibliographystyle{IEEEtran}
\bibliography{IEEEfull,reference}
\ifCLASSOPTIONcaptionsoff
  \newpage
\fi




\end{document}